# Biological and Radiological Dictionary of Radiomics Features: Addressing Understandable AI Issues in Personalized Prostate Cancer; Dictionary Version PM1.0

Mohammad R. Salmanpour[1,2,3*], Sajad Amiri[4], Sara Gharibi[4], Ahmad Shariftabrizi[5], Yixi Xu[3], William B Weeks[3], Arman Rahmim[1,2], Ilker Hacihaliloglu[1,6]

[1]Department of Radiology, University of British Columbia, Vancouver, BC, Canada
[2]Department of Integrative Oncology, BC Cancer Research Institute, Vancouver, BC, Canada
[3]AI for Good Research Lab, Microsoft Corporation, Redmond, WA, US
[4]Technological Virtual Collaboration (TECVICO Corp.), Vancouver, BC, Canada
[5]Department of Radiology, University of Iowa Carver College of Medicine, IA, US
[6]Department of Medicine, University of British Columbia, Vancouver, BC, Canada

(*) Corresponding Author:
Mohammad R. Salmanpour, PhD
Department of Radiology, University of British Columbia, Vancouver, BC, Canada
Email: *m.salmanpour@ubc.ca*

**ABSTRACT**
**Background:** Artificial intelligence (AI) can advance medical diagnostics, but interpretability limits its clinical use. This work links standardized quantitative imaging features (radiomics features, RF) extracted from medical images with clinical frameworks like PI-RADS, ensuring AI models are understandable and aligned with clinical practice.
**Methods:** We investigate the connection between visual semantic features defined in PI-RADS and associated risk factors, moving beyond abnormal imaging findings, establishing a shared framework between medical and AI professionals by creating a standardized dictionary of biological/radiological RFs. Subsequently, 6 interpretable and seven complex classifiers, linked with nine interpretable feature selection algorithms (FSA) applied to risk factors, were extracted from segmented lesions in T2-weighted imaging (T2WI), diffusion-weighted imaging (DWI), and apparent diffusion coefficient (ADC) multiparametric-prostate MRI sequences to predict the UCLA scores. We then utilized the created dictionary to interpret the best-predictive models.
**Results:** Combining T2WI, DWI, and ADC with FSAs including ANOVA F-test, Correlation Coefficient, and Fisher Score, and utilizing logistic regression, identified key features: The 90th percentile from T2WI, which captures hypo-intensity related to prostate cancer risk; Variance from T2WI, indicating lesion heterogeneity; shape metrics including Least Axis Length and Surface Area to Volume ratio from ADC, describing lesion shape and compactness; and Run Entropy from ADC, reflecting texture consistency. This approach achieved the highest average accuracy of 0.78±0.01, significantly outperforming single-sequence methods (p-value<0.05).
**Conclusion:** The developed dictionary for Prostate-MRI (PM1.0) serves as a common language, fosters collaboration between clinical professionals and AI developers to advance trustworthy AI solutions that support reliable/interpretable clinical decisions.

**Keywords:** Dictionary of Radiomics Features; Understandable AI; Prostate Cancer; Prostate Imaging Reporting and Data System (PI-RADS).

## 1. INTRODUCTION

In the rapidly evolving landscape of medical technology, integration of Artificial Intelligence (AI) has the potential to revolutionize diagnostics, prognostication, and treatment planning. However, adoption of AI in healthcare is often hindered by concerns regarding transparency and trustworthiness, primarily due to the "black box" nature of many advanced AI models, such as deep learning systems [1, 2]. To address these concerns, explainable AI (XAI) and interpretable AI (IAI) have emerged as critical approaches aimed at enhancing the transparency and reliability of AI systems in medical settings [3, 4, 1, 5]. XAI focuses on providing post-hoc explanations for complex models through methods like feature importance scores, Shapley Additive Explanations (SHAP), and heatmaps, which help elucidate AI-driven diagnoses and treatment recommendations [6, 7]. Conversely, IAI leverages inherently transparent models, such as decision trees, which offer clear, rule-based decision-making processes [8]. Use of XAI and IAI is essential in healthcare, where AI-driven decisions can significantly impact patient outcomes, ensuring that clinical professionals can trust and effectively integrate AI insights into their practice [9, 10, 11].

An important frontier in image-guided interventions is the use of radiomics features (RF), which extract extensive quantitative data from medical images to uncover disease characteristics not discernible to the human eye, such as tumor heterogeneity and genetic profiles [12, 13, 14, 15, 16]. These high-dimensional data facilitate personalized medical care and improve the integration of RFs into clinical workflows, enhancing predictive assessments and patient outcomes [17, 18]. RFs have shown substantial promise in identifying cancer biomarkers, enabling early detection, refining prognostic models [19, 20, 21], and advancing the understanding of disease mechanisms for targeted therapies [22]. Of potentially great value to standardization and interpretability of RFs is Prostate Imaging Reporting and Data System (PI-RADS), which provides a unified framework for interpreting prostate MRI findings [23, 24, 25, 26, 27, 28]. Since its introduction in 2012, PI-RADS has undergone several updates (v2 in 2015 and v2.1 in 2019) to enhance consistency and accuracy in prostate cancer detection [28, 23, 24, 25]. PI-RADS standardizes prostate MRI interpretation along with enhancing the explainability and interpretability of RFs. This framework improves diagnostic accuracy, ensures reliable RFs extraction across studies and institutions [26], and boosts the reproducibility and clinical relevance of radiomic research [27].

PI-RADS integrates multiple MRI sequences—includingT2-weighted imaging (T2WI), diffusion-weighted imaging (DWI), apparent diffusion coefficient (ADC) and dynamic contrast-enhanced imaging (DCE)—to provide comprehensive functional and anatomical insights that improve prostate cancer detection and characterization [23, 29, 25, 30]. For instance, T2WI identifies structural abnormalities, DWI assesses tissue cellular density through apparent diffusion coefficient (ADC) values, and DCE highlights tissue vascularity [23, 24, 25, 30]. Correlating RFs with PI-RADS elements can enable the creation of transparent models that link imaging data with clinically understood criteria, thereby enhancing the trustworthiness and predictive power of AI-driven diagnostic tools [2, 31, 32, 33].

Moreover, interpreting RFs addresses key challenges in both XAI and IAI by offering transparent insights into AI model decision-making processes, fostering trust by clinical professionals [12, 34, 35]. Standardization efforts, such as those by the Image Biomarker Standardization Initiative (IBSI), ensure the reproducibility of RFs, supporting their clinical adoption and integration with clinical and genomic data for personalized disease treatment [14, 13, 36, 16, 37, 38, 39, 40]. Despite significant advancements, there remains a need for a more comprehensive and clinically actionable understanding of RFs [41, 14, 42, 43, 44]. Recent studies emphasize the integration of biological and radiological data to enhance the interpretability and relevance of RFs, thereby improving the explainability and trust in AI models used in clinical settings [13, 45, 42, 43, 46, 47]. IAI models are thus essential for medical diagnosis and biological data analysis, enabling a clear understanding of the decision-making processes underlying AI predictions [48, 23]. Balancing interpretability with accuracy remains a primary focus in model development, with rule-based methods and personalized interpretable classifications offering promising solutions [49, 50, 51, 52, 53, 54, 55, 56]. Additionally, effective feature selection plays a critical role in enhancing model performance by reducing complexity and improving interpretability, which is particularly important in medical applications where understanding the basis of predictions is crucial for clinical decision-making and trust [57, 58, 59, 60, 61, 62, 63, 64, 65, 66].

This study explores a unified approach combining a relationship-based RF dictionary and a classification framework to advance prostate cancer diagnosis and management. The dictionary establishes connections between PI-RADS visual semantic features and quantitative imaging data, improving the explainability and clinical utility of AI models. By translating descriptive clinical language into actionable data, this method bridges the gap between clinical professional expertise and computational insights. Leveraging RFs extracted from T2WI, DWI, and ADC sequences, the classification pipeline employs both interpretable and complex algorithms, optimized through grid search and rigorously validated via cross-validation and external testing. Selected features are presented transparently to enhance model interpretability, enabling clinicians to trust and understand the decision-making process. This comprehensive framework not only enhances diagnostic accuracy but also fosters personalized treatment planning, offering a significant step toward better patient outcomes in prostate cancer care.

Tomaszewski and Gillies [12] preliminarily studied the reintroduction of biological meaning into radiomics through approaches such as genomic correlations, microscopic pathologic textures, and histopathologic markers. They emphasized integrating these analyses during model development or validation to uncover underlying biological mechanisms. Their work highlights the need to standardize these practices to enhance radiomics' clinical and scientific impact. Other related work and contributions include those by Gillies et al. [41] and Aerts et al. [14], who described the transformation of medical images into data and linked RFs to tumor phenotypes, yet no comprehensive dictionary linking RFs to biological and radiological data is available yet [67, 17, 19]. Stoyanova et al. [68] and Gnep et al. [69] demonstrated the integration of RFs with genomic data and clinical relevance in prostate cancer [69, 70]. PI-RADS, developed by Weinreb and Barentsz [23], standardizes prostate MRI interpretations, while the UCSF-CAPRA score by Cooperberg and Carroll [71] integrates clinical features for outcome prediction. Rudin [2] highlighted that a comprehensive RFs dictionary is essential for interpretable AI in critical decisions like cancer diagnosis, enhancing AI transparency and clinical relevance in XAI. Papanikolaou et al. [72] focused on preventing methodological errors and addressing key workflow challenges in developing clinically meaningful RFs. Mercaldo et al. [73] used GradCAM and SmoothGrad to visually explain convolutional neural networks (CNN) decisions and highlight key image regions in lung cancer, enhancing interpretability. Vimbi et al. [74] concluded that methods like Local Interpretable Model-agnostic Explanations (LIME) and SHAP offer local and global explanations, address model opacity and bias, thereby enhancing the trustworthiness of AI models in clinical settings.

Timmeren [75] highlighted the limitations of RFs analysis and emphasized creating a detailed RFs dictionary linked to biological and radiological meanings. This bridges computational methods and clinical applications, enhancing the accuracy of AI-driven cancer detection and improving patient care. Techniques like GradCAM and SmoothGrad are employed to enhance model transparency in cancer detection but lack a standardized RFs dictionary [76]. While techniques like LIME, SHAP, and neural network attention are widely used to highlight key input features influencing model predictions, they do not address the creation of a detailed RFs dictionary [74]. Techniques like LIME, SHAP, and neural network attention highlight key input features influencing model predictions. For example, [77] introduced SHAP in clinical settings to provide accurate feature attributions for complex models, while Lundberg et al. [78] demonstrated that explainable AI makes CNN decision processes transparent for diagnosing diabetic retinopathy. Thus, a significant need remains for developing a comprehensive biological and radiological dictionary of RFs to enhance AI interpretability and clinical applicability in medical imaging, which the present work attempts. In the following, we first utilize visual semantic feature definitions derived from the PI-RADS scoring system to explore conceptual biological and radiological relationships between imaging RFs and visual semantic features. Next, we apply ML techniques, including interpretable feature selection algorithms (FSA) combined with a mix of interpretable and complex classifiers, to predict prostate cancer risk outcomes. Finally, we leverage the constructed dictionary to demonstrate why the selected features, in combination with classifiers, achieved high performance and validated the clinical relevance of the RFs.

## 2. MATERIAL AND METHOD

### 2.1. Exploring Relationship Between RFs and PI-RADS Scoring System

We utilized PyRadiomics [79], standardized in reference to the IBSI [16], to extract a total of 119 standardized RFs from each T2WI, DWI, ADC MRI sequences, including 19 First Order Features (FO), 16 3D Shape Features (SF), 10 2D SF, 23 Gray Level Co-occurrence Matrix Features (GLCM), 16 Gray Level Size Zone Matrix Features (GLSZM), 16 Gray Level Run Length Matrix Features (GLRLM), 5 Neighboring Gray Tone Difference Matrix Features (NGTDM), and 14 Gray Level Dependence Matrix Features (GLDM). A comprehensive list of RFs is provided in Supplemental Table S1.

**PI-RADS Scoring.** The criteria for PI-RADS scoring in the different prostate zones, as outlined in Supplemental Table S2, and in accordance with the uploaded PI-RADS v2.1 guidelines. Figure 1 (Middle) illustrates the PI-RADS score calculation in both PZ and TZ. Furthermore, comprehensive description of semantic visual assessment features is outlined in Supplemental Table S3.

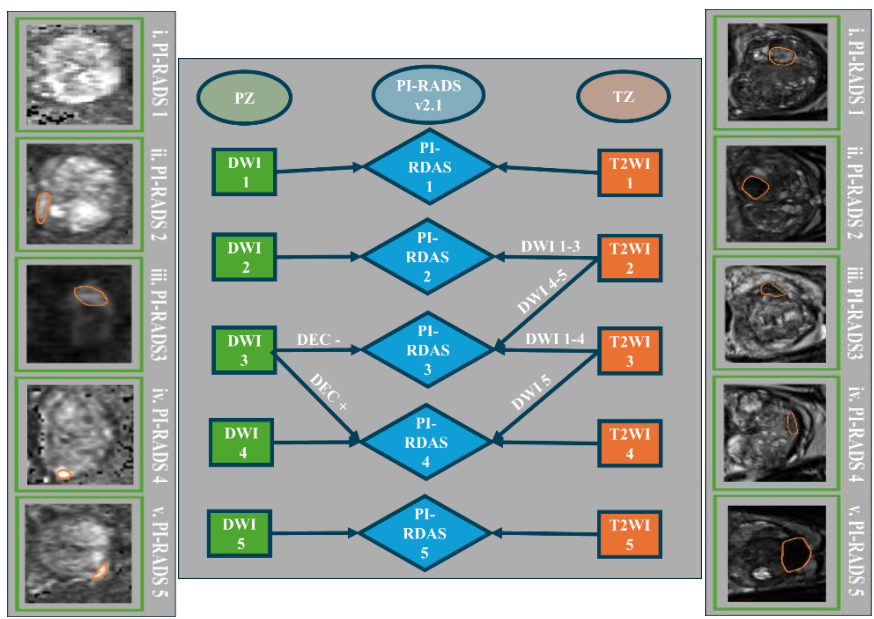

**Fig. 1.** (Middle) PI-RADS v2.1 (Prostate Imaging–Reporting and Data System) is a structured reporting system for multiparametric prostate MRI, used to assess prostate cancer. (Left) Examples of PI-RADS scores 1 to 5 in the Peripheral Zone (PZ). (Right) Examples of PI-RADS scores 1 to 5 in the Transition Zone (TZ). DWI: Diffusion-weighted imaging, DCE: Dynamic contrast-enhanced imaging, T2WI: T2-weighted imaging.

**Transition Zone (TZ).** TZ, as illustrated in Supplemental Figure S1, is primarily assessed using T2WI, with scores assigned in Figure 1 as follows. A score of 1 on T2WI indicates a normal appearing TZ, which is rare and shows homogeneous intermediate signal intensity (see Figure 1, Right, i). It may also present as a round, completely encapsulated nodule, known as a "typical nodule," which is typical of benign conditions like benign prostatic hyperplasia. This score corresponds to a PI-RAD score of 1. A score of 2 on T2WI indicates the presence of a mostly encapsulated nodule, a homogeneous circumscribed nodule without encapsulation, or a homogeneous mildly hypo-intense area between nodules (see Figure 1, Right, ii). If the DWI score of the lesion is between 1-3, the PI-RAD score is 2, otherwhile the PI-RAD score is 3. A score of 3 on T2WI indicates heterogeneous signal intensity with obscured margins, including others that do not qualify as 2, 4, or 5 (see Figure 1, Right, iii). When a lesion receives a T2WI score of 3, DWI is used as a supplementary tool to further assess the likelihood of malignancy. If the DWI score is between 1 and 4, the PI-RADS score remains 3, indicating a lower likelihood of significant cancer. However, if the DWI score is 5, the PI-RADS score increases to 4, suggesting a higher suspicion of malignancy [23]. A score of 4 on T2WI indicates a lenticular or non-circumscribed, homogeneous, moderately hypo-intense area that is suspicious for malignancy (see Figure 1, Right, iv). Finally, a score of 5 on T2WI indicates the same characteristics as a score of 4, but the lesion is equal to or greater than 1.5 cm in greatest dimension or shows definite extra-prostatic extension/invasive behavior, making it highly suspicious for malignancy (see Figure 1, Right, v).

**Peripheral Zone (PZ).** PZ, as illustrated in Supplemental Figure S1, is primarily assessed using DWI, with scores assigned in Figure 1 as follows: Score 1 indicates no abnormalities on ADC and high b-value DWI (between 1400 and 2000 s/mm²), corresponding to PI-RADS 1 (see Figure 1, Left, i). Score 2 is characterized by linear or wedge-shaped hypo-intensity on ADC and/or linear or wedge-shaped hyper-intensity on high b-value DWI, aligning with PI-RADS 2 (see Figure 1, Left ii). Score 3 denotes focal hypo-intensity on ADC that is distinct from the background and/or focal hyper-intensity on high b-value DWI. In this case, the ADC may be markedly hypo-intense, or the high b-value DWI may be markedly hyper-intense, but not both (see Figure 1, Left iii). To determine the PI-RADS in this scenario, DCE sequence must be evaluated: if negative, it corresponds to PI-RADS 3, while a positive result indicates PI-RADS 4. Score 4 is defined by focal marked hypo-intensity on ADC and marked hyper-intensity on high b-value DWI, which is equivalent to PI-RADS 4 (see Figure 1, Left iv). Score 5 mirrors Score 4 but involves lesions measuring ≥1.5 cm in greatest dimension or exhibiting definite extra-prostatic extension or invasive behavior corresponding to PI-RADS 5 (see Figure 1, Left v). For the assessment of DCE, it serves to clarify ambiguous DWI findings: negative DCE indicates no early and intense enhancement, suggesting a lower likelihood of significant cancer, whereas positive DCE indicates early and intense enhancement, which implies the presence of clinically significant cancer.

**Procedure for Creating a Relationship-Based Dictionary.** This study investigates the relationship of RFs associated with different PI-RADS scores focusing on deviations from normal imaging. Typically, PZ of a normal prostate exhibits uniform pixel intensity values, indicating low randomness and resulting in a lower standard deviation in c features. However, in PZ-based PI-RADS 3, the imaging on DWI sequences becomes heterogeneous. This increased heterogeneity leads to a higher standard deviation in FO features compared to normal tissue. This method by establishing meaningful relationships between the semantic visual assessment features outlined in PI-RADS and related RFs. By focusing on the semantic meaning of RFs and visual assessment features derived from PI-RADS scoring system, we have developed a conceptual method to map the descriptive language as used by radiologists and clinicians to the quantitative data extracted from imaging studies. This approach not only enhances the explainability and interpretability of AI models by addressing the relationship between RFs and model outputs, but also improves the clinical relevance and usability of RFs in practice [12, 80, 81]. Furthermore, this dictionary was validated by two medical physicists, and four medical professionals—including two doctors of medicine, a radiologist, and a physician whom are familiar with RFs and AI analysis.

### 2.2. Understandable Classification Task
**Patient Data.** We utilized a dataset of 1080 prostate cancer patients from The Cancer Imaging Archive [82], which included ultrasound, different sequences of MRI images (T2WI, DWI, and ADC), and corresponding segmented PZ or/and TZ cancerous lesions. This dataset originates from biopsy sessions conducted with the Artemis biopsy system integrated with MRI, focusing on the University of California, Los Angeles (UCLA) prostate cancer index. Patients with suspected prostate cancer, indicated by elevated Prostate-Specific Antigen levels and/or suspicious imaging findings, were consecutively enrolled. Eligible participants included those who consented and either underwent or were scheduled for a routine, standard-of-care prostate biopsy at the UCLA Clark Urology Center. The UCLA scoring system, launched in 2010 and mirroring the ESUR PI-RADS [28], categorizes MRI findings on a scale from 1 to 5. This scale delineates 1 as very low (seems normal), 2 as mildly abnormal, 3 as moderately abnormal, 4 as highly abnormal, and 5 as profoundly abnormal in terms of cancer suspicion levels [83]. In this investigation, scores from 1 to 3 were grouped under the low-risk category, while scores from 4 to 5 were categorized as high-risk. High patient participation highlights their commitment to addressing factors that affect their quality of life [84]. Imaging was conducted using high-resolution ultrasound devices (Hitachi Hi-Vision 5500 7.5 MHz or Noblus C41V 2-10 MHz probes) and 3 Tesla Siemens MRI scanners (Trio, Verio, or Skyra), catering to patients with elevated PSA levels or suspicious imaging findings and utilizing various imaging protocols and technical specifications.

**Interpretable and Complex Machine Learning (ML) Algorithms.** Interpretable machine learning (ML) or IAI algorithms are designed to ensure that models are transparent [85]. ML systems are increasingly influencing societal decisions, yet many models operate as black boxes, obscuring the logic behind their predictions [86]. This lack of transparency is particularly problematic in regulated domains where auditability is essential. As a result, there's a growing demand for interpretable ML algorithms that not only provide accurate predictions but also make their decision-making processes understandable [53, 54, 56]. This section examines the importance of these algorithms and the methods developed to enhance interpretability, along with the challenges in assessing explanation quality [87, 88, 89, 90].

Fig. 2 Depicts schematic diagram of the proposed workflow:

i) Cancerous lesion masks were delineated by experts on MRI sequences, including T2WI, DWI, and ADC images, to ensure accurate segmentation of the regions of interest (see supplemental Figure S2);

ii) The delineated masks were independently reviewed and verified by a second expert to ensure consistency and reliability of the segmentations;

iii) Images were normalized using a min/max normalization function to standardize intensity ranges across different imaging modalities and enhance feature extraction;

iv) RFs were extracted from the masks applied to the preprocessed MRI sequences using the standardized Pyradiomics software package, ensuring robust feature computation;

v) Patient datasets were split into 85% for five-fold cross-validation and 15% for external nested testing, with the external set evaluated in each fold to assess generalizability;

vi) The extracted RFs were normalized by min/max function using training set parameters to minimize data bias and maintain consistency across folds;

vii) Nine interpretable FSAs, such as the Chi-Square Test (CST), Correlation Coefficient (CCF), Mutual Information (MIS), Variance Threshold (VTS), ANOVA F-test (AFT), Information Gain (IGS), Univariate Feature Selection (UFS), Fisher Score (FSF) [91, 92], and Least Absolute Shrinkage and Selection Operator (LAS) [93, 94], were applied to the training datasets to identify the most relevant subsets of features for classification tasks;

viii) A combination of 13 classification algorithms (CA), including six interpretable classification models, such as Decision Tree Classification (DTC) [86] , 2) Logistic Regression (LOC) [95], Linear Discriminant Analysis (LDA) [96], Naive Bayes Classifier (NBC), K-Nearest Neighbors (KNN), RuleFit Classifier (RUC) [29], along with seven black-box ensemble techniques like Random Forest (RFC) [97], XGBoost Classifier (XGB), LightGBM Classifier (LGB), CatBoost Classifier (CBC) [98], Support Vector Machine (SVM) [99], Stacking Classifier (STC), and Multilayer Perceptron (MLP) [100] were applied to the selected features; and

ix) The top five features identified by FSAs, based on classification performance, were selected and incorporated into a dictionary (Table 1) to provide interpretability and clinical relevance for the chosen imaging features. The UCLA scores utilized a target variable and RFs as inputs for CAs. All algorithms are elaborate in supplemental section 1.2. Using this dictionary enhances interpretability and explainability, allowing clinicians to understand model decisions and how features influence prostate cancer classification. For AI developers, it provides clear insights into feature importance, enabling more transparent model design, improved performance, and alignment with clinical needs.

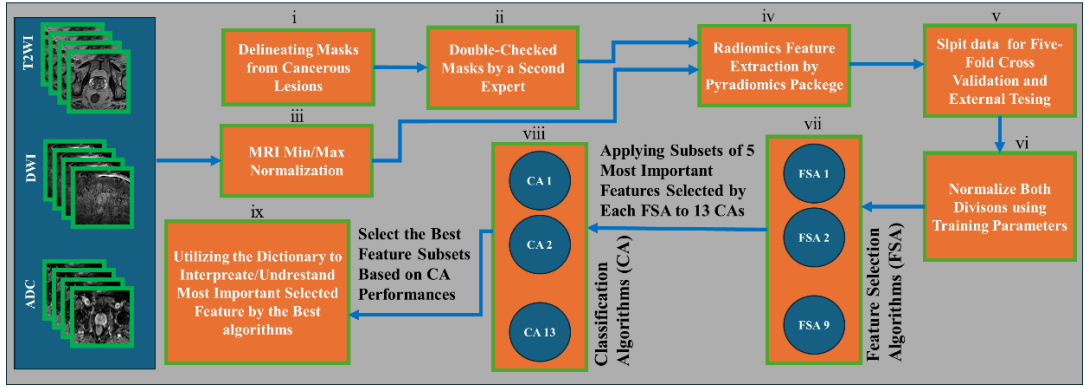

**Fig. 2.** Schematic diagram of the proposed workflow: i) Delineating cancerous lesion masks by experts, ii) Verifying delineated masks through a second expert review, iii) Normalizing images using a Min/Max normalization function, iv) Extracting radiomics features from masks applied to preprocessed MRI sequences using Pyradiomics software, v) Splitting datasets for five-fold cross-validation and external nested testing, vi) Normalizing radiomics features based on training parameters, vii) Applying 9 feature selection algorithms (FSA) to the training set of different datasets to select the best subset of features, viii) Applying 13 interpretable and complex classification algorithms (CA) to the features selected by FSAs, and ix) Selecting the top five features identified by FSAs based on classification performance and utilizing a created dictionary to explain the selected imaging features.

## 3. RESULTS

### 3.1. Radiological/Biological Definition of RFs for Addressing Understandable AI (UAI) Issues

Understandable AI (UAI) encompasses both IAI and XAI, aiming to bridge technical transparency with domain-specific clarity. While XAI/IAI focus on revealing how models work or make decisions, UAI extends this by aligning outputs with domain standards, ensuring both technical and practical understanding by experts and stakeholders. This study analyzed the PI-RADS scoring system, which evaluates prostate lesions based on multiparametric MRI, linking its visual semantic features to RFs such as FO, SF, and matrix-based features. By developing a dictionary that connects PI-RADS visual semantic features to RFs, the study enhances the quantitative basis for semantic descriptions, improving feature selection related to UAI. Unlike methods like SHAP and LIME, this approach provides clearer explanations for the relationships between features and outcomes, reducing the "black box" nature of AI models and offering physicians greater transparency and trust in their decision-making processes. A comprehensive list of relationships between PI-RADS elements and RFs is provided in Supplemental Table S4.

### 3.1.1. PI-RADS 1&2

A PI-RADS 1 in the TZ typically indicates a normal appearance on T2WI, where the lesion may exhibit a homogeneous intermediate signal intensity or appear as a round, well-encapsulated nodule. The visual semantic feature of homogeneity is quantified by various RFs. PI-RADS 2 in TZ generally indicates a mildly hypo-intense lesion on T2WI with a corresponding DWI score of 1 to 3. A T2WI score of 2 refers to a mildly hypo-intense lesion, which may appear as an encapsulated or atypical nodule. DWI score 1 indicates no abnormality on ADC and high b-value DWI while DWI score 2 refers to linear or wedge-shaped hypo-intensity on ADC and/or linear or wedge-shaped hyper-intensity on high b-value DWI. DWI score 3 describes focal hypo-intensity on ADC and/or focal hyper-intensity on high b-value DWI, with the possibility of being markedly hypo-intense on ADC or markedly hyper-intense on high b-value DWI, but not applicable to both.

**Heterogeneity and Uniformity Visual Semantic Features.** In multiparametric prostate MRI, homogeneity, heterogeneity, and uniformity describe signal intensity patterns. Homogeneity indicates consistent texture, brightness, and contrast across a lesion, while heterogeneity highlights variability, with fine heterogeneity showing small irregularities and coarse heterogeneity reflecting larger patterns. Uniformity emphasizes even signal distribution, complementing homogeneity by focusing on equal intensity across the lesion. These are key features for interpreting prostate cancer lesions. RFs like Large Dependence Emphasis and Large Dependence High Gray Level Emphasis from the GLDM category (GLDM_LDHGLE and GLDM_LDE), highlight homogeneity in high values and coarse heterogeneity in low values, focusing on larger structures. Conversely, RFs like Small Area Low Gray Level Emphasis from the GLSZM category (GLSZM_SAHGLE) and Short Run Low Gray Level Emphasis from the GLRLM category (GLRLM_SRLGLE) emphasize fine heterogeneity, capturing smaller, non-uniform structures. Additionally, general features like Standard Deviation from the FO category (FO_SD) represent overall heterogeneity by measuring the variation or dispersion in intensity values across the texture. By contrast, heterogeneity refers to a structure with dissimilar components or elements, appearing irregular or variegated. As shown in Table 1 (second row), homogeneity or heterogeneity visual semantic features and the associated RFs serve as tools to evaluate PI-RADS 1 lesions, as follows:

- Entropy from the FO category (FO_En) measures randomness in image values, where lower entropy signifies more consistency and less variation.
- FO_SD measures the amount of variation or dispersion from the average intensity values in an image. By definition, a higher value indicates greater differences between intensities, leading to more heterogeneity and less uniformity. This principle applies similarly to Variance (FO_V), Mean Absolute Deviation (FO_MAD), and Robust Mean Absolute Deviation (FO_rMAD) and Root Mean Squared (FO_RMS) from the FO category.
- Kurtosis from FO category (FO_Ku) measures the 'peakedness' of intensity distributions, with lower kurtosis indicating a more uniform intensity distribution.
- Range from the FO category (FO_R) represents the intensity variation, where a lower range indicates less difference between intensities and greater homogeneity. However, this does not always hold true in the reverse case.
- Inverse Difference Moment feature from the GLCM category (GLCM_IDM), also known as homogeneity, measures local homogeneity by assessing the similarity of intensity values between neighboring voxels, with high GLCM_IDM values supporting the homogeneous nature of PI-RADS 1 lesions. Inverse Difference Moment Normalized feature from the GLCM category (GLCM_IDMN) is the normalized version of GLCM_IDM.
- Joint Entropy from the GLCM category (GLCM_JEn) measures randomness or variability in the distribution of neighboring intensity values, reflecting the heterogeneity of the texture.
- The Uniformity feature from the FO category (FO_Un) calculates the sum of the squares of intensity values, where greater uniformity suggests higher homogeneity, while Skewness from the FO category (FO_Sk) measures the asymmetry of the intensity distribution, with values near zero indicating symmetry, which is characteristic of homogeneous tissues.
- Inverse Difference from the GLCM category (GLCM_ID) measure of the local homogeneity of an image. With more uniform gray-levels, the denominator will remain low, resulting in a higher overall value. The normalized version of the GLCM_ID is called Inverse Difference Normalized (GLCM_IDN).
- Joint Energy from the GLCM category (GLCM_JE) measures the frequency and uniformity of co-occurring intensity pairs in an image. Higher values indicate a greater frequency of neighboring intensity pairs, reflecting homogenous and consistent, uniform patterns in the texture. These values are often associated with regions where specific gray-level intensity pairs are repeatedly observed. Similarly, Cluster Tendency from the GLCM category (GLCM_CT) assesses the grouping of voxels based on gray-level similarities, with higher values indicating increased uniformity and more cohesiveness in voxel clusters.
- Correlation from the GLCM category (GLCM_Corr) evaluates the linear dependency between neighboring voxel intensities, with high values indicating predictable relationships.
- Contrast from the GLCM category (GLCM_Co) quantifies local intensity variation, emphasizing differences away from the diagonal (where i=j). Higher values indicate greater disparity in intensity among neighboring voxels, reflecting lower homogeneity in the image texture.
- Cluster Prominence and Cluster Shade from the GLCM category (GLCM_CP and GLCM_CS) assess asymmetry and skewness, where lower values are consistent with homogeneity.
- Autocorrelation from the GLCM category (GLCM_AC) measures the degree of fineness and coarseness in texture. Greater coarseness results in reduced uniformity and homogeneity, leading to increased heterogeneity. Difference Entropy (GLCM_DiEn) from the GLCM category measures the randomness or variability in the intensity differences between neighboring pixels. This metric reflects the degree of heterogeneity in the texture; high values indicate greater randomness and complexity in intensity differences, signifying a heterogeneous texture with diverse pixel values and intricate spatial relationships.
- The Informational Measure of Correlation features from the GLCM category (GLCM_IMC1 and GLCM_IMC2) assess the complexity and correlation of gray-level distributions, with higher values suggesting higher homogeneity.
- Maximal Correlation Coefficient from the GLCM category (GLCM_MCC) measures texture complexity, where higher values indicate more uniform and predictable textures.

- Difference Average from the GLCM category (GLCM_DA) measures the relationship between occurrences of pairs with similar intensity values and occurrences of pairs with differing intensity values. A low value suggests that many pairs of data points have similar intensity values, indicating a more homogeneous dataset.
- Sum Entropy from the GLCM category (GLCM_SEn) is a summation of neighborhood intensity value differences. The feature quantifies the diversity and distribution of pixel intensity relationships in an image. A high value indicates a wide variety of intensity levels and relationships, reflecting a more complex and heterogeneous texture.
- Sum of Squares from the GLCM category (GLCM_SQ) measures the variance in the distribution of neighboring intensity level pairs relative to the mean intensity in the GLCM. Lower values indicate that pixel intensities are more similar, suggesting a homogeneous texture, typical of less complex tissues.
- Maximum Probability from the GLCM category (GLCM_MP) represents the occurrence of the most frequent pair of neighboring intensity values. A high value indicates a dominant intensity pair that recurs frequently, suggesting homogeneity or regularity in the texture, often observed in uniform tissues or structures.
- Inverse Variance from the GLCM category (GLCM_IV) assesses how the intensity of pixel values in an image varies. A low value indicates that pixel values are closely packed together, which suggests homogeneity.
- Dependence Non-Uniformity from the GLDM category (GLDM_DN) measures similarity of dependence throughout the image, with a lower value indicating more homogeneity among dependencies in the image. The normalized version of GLDM_DN is called Dependence Non-Uniformity Normalized (GLDM_DNN).
- Dependence Variance from the GLDM category (GLDM_DV) measures the variance in dependence size within the image. Higher values indicate greater differences in zone sizes, leading to increased heterogeneity.
- Gray Level Non-Uniformity from the GLDM category (GLDM_GLN) measures the similarity of gray-level intensity values within the image. A lower value indicates greater similarity in intensity values, reflecting a more uniform distribution of pixel intensities and suggesting higher homogeneity.
- Dependence Entropy from the GLDM category (GLDM_DeEn) quantifies the spatial relationships of gray-levels by measuring the distribution of connected voxels (dependencies) within a specified distance threshold that share the same gray-level intensity. It specifically evaluates the randomness or complexity (heterogeneity) in the distribution of these dependencies, providing insight into the texture variability and structural irregularities in the image.
- Gray Level Variance from the GLDM category (GLDM_GLV) measures the variation in gray-levels within an image. Higher GLDM_GLV values indicate greater differences between intensities around a specific mean, reflecting increased heterogeneity in the texture.
- Dependence Variance from the GLDM category (GLDM_DV) measures the variance in dependence size within the image. A higher value indicates more diverse relationships between different gray-levels (or pixel values), reflecting a broader range of spatial relationships and a more complex texture pattern.
- GLDM_LDE measures the distribution of large dependencies within the image. Higher values indicate larger dependencies and more homogeneous textures. This metric is valuable for analyzing and characterizing texture, as it highlights the presence of large, homogeneous dependencies.
- Small Dependence Low Gray Level Emphasis from the GLDM category (GLDM_SDLGLE) measures the joint distribution of small dependencies with lower gray-level values. This texture feature evaluates the frequency of small, low-intensity gray-levels occurring together in an image, reflecting the uniformity or consistency of the tissue texture.
- Small Dependence High Gray Level Emphasis from the GLDM category (GLDM_SDHGLE) is similar to GLDM_SDLGLE but focuses on the joint distribution of small dependencies with higher gray-level values.
- Small Dependence Emphasis from the GLDM category (GLDM_SDE) is a measure of the distribution of small dependencies, where a higher value indicates the presence of smaller, less homogeneous regions in the image texture, contributing to more texture heterogeneity.
- Gray Level Non-Uniformity from the GLSZM category (GLSZM_GLN) measures the variability of gray-level intensity values in an image. A lower value indicates greater homogeneity in intensity values, resulting in a more homogeneous image. Gray Level Non-Uniformity Normalized (GLSZM_GLNN) is the normalized version of GLN.
- Size-Zone Non-Uniformity (GLSZM_SZN) from the GLSZM category measures the variability of size zone volumes, with lower values indicating more uniform zones. The normalized version of GLSZM_SZN is referred to as GLSZM_SZNN.
- Zone Percentage from the GLSZM category (GLSZM_ZP) measures the coarseness of the texture by taking the ratio of number of zones and number of voxels in the ROI. Higher values indicate that a larger portion of the

- ROI consists of smaller zones. This suggests a finer texture and implies more intricate variations within the image, which is a hallmark of heterogeneity.
- Small Area Emphasis from the GLSZM category (GLSZM_SAE) measures the distribution of small-sized zones. Higher values indicate a greater emphasis on smaller zones, reflecting finer textures and more intricate variations in intensity. This typically suggests a more complex and heterogeneous tissue structure.
- Small Area Low Gray Level Emphasis from the GLSZM category (GLSZM_SALGLE) measures the proportion of smaller size zones with lower gray-level values in an image. High values indicate fine, heterogeneous textures, often associated with hypo-intense lesions.
- Small Area High Gray Level Emphasis from the GLSZM category (GLSZM_SAHGLE) measures the proportion of smaller size zones with higher gray-level values in an image. High values indicate fine, heterogeneous textures, often associated with hyper-intense lesions.
- Additionally, Zone Entropy from the GLSZM category (GLSZM_ZE) measures the uncertainty or randomness in the distribution of zone sizes and gray-levels, where higher values indicate more heterogeneity in the texture patterns.
- Gray Level Variance from the GLSZM category (GLSZM_GLV) measures the variance in gray-level intensities across the zones. Higher variance values indicate greater differences in intensity, reflecting increased heterogeneity within the image.
- Large Area Emphasis (GLSZM_LAE) from the GLSZM category measures the distribution of large area size zones within an image. A higher value indicates a predominance of larger zones, suggesting coarser textures and greater homogeneity in the image's overall texture.
- Zone Variance from the GLSZM category (GLSZM_ZV) measures the variance in zone-size volumes for the zones. High values indicate a wide range of zone-sizes, suggesting a heterogeneous texture.
- Short Run High Gray Level Emphasis from the GLRLM category (GLRLM_SRHGLE) measures the joint distribution of short run lengths and higher gray-level values. Higher values suggest areas with fine heterogeneous textures, characterized by short runs and elevated intensity values.
- Short Run Emphasis from the GLRLM category (GLRLM_SRE) measures the distribution of short run lengths, where higher values indicate shorter run lengths and finer textural details. Images with high values are characterized by numerous short runs, reflecting greater variability in intensity values. This variability contributes to a heterogeneous texture with fine details, often described as a fine, heterogeneous texture.
- Long Run Emphasis from the GLRLM category (GLRLM_LRE) measures the distribution of long run lengths, where higher values indicate longer run lengths and coarser structural textures. Images with high values are typically characterized by larger, more consistent particles, reflecting a coarse texture. Very high values of this feature suggest an even more consistent and uniform texture, indicative of a high level of homogeneity. This homogeneity arises from reduced variability in intensity values across the image, creating a stable and predictable pattern that defines the coarse, structured nature of the texture.
- Run Length Non-Uniformity from the category GLRLM (GLRLM_RLN) measures the similarity of run lengths throughout the image, with a lower value indicating more homogeneity among run lengths in the image. A low value indicates that the run lengths are more uniform across the image. This suggests fewer variations in pixel intensity or color, resulting in areas of the image that exhibit similar characteristics. In other words, when the run lengths are consistent and uniform, it reflects a more homogenous texture or pattern in the image. Run Length Non-Uniformity Normalized (GLRLM_RLNN) is the normalized version of GLRLM_RLN with the same representation.
- GLRLM_SRLGLE measures the joint distribution of short run lengths with low gray-level values. High values indicate the presence of fine-heterogeneous, hypo-intense structures in the texture.
- Run Entropy from the GLRLM category (GLRLM_REn) quantifies uncertainty or randomness in the distribution of run lengths and gray-levels within a texture. A high value indicates greater heterogeneity, reflecting a diverse and complex texture with significant variation in patterns and gray-levels
- Gray Level Non-Uniformity from the GLRLM category (GLRLM_GLN) measures the similarity of gray-level intensity values in an image. A lower value indicates greater similarity in intensity values, conveying more homogeneity. The normalized version of GLRLM_GLN is referred to as GLRLM_GLNN.
- Gray Level Variance from the GLRLM category (GLRLM_GLV) measures the variability in gray-level intensities across the runs. Higher values indicate greater variance in gray-levels, suggesting a more heterogeneous texture.
- Run Variance from the GLRLM category (GLRLM_RV) measures the variance in run lengths. A high value indicates significant variation in the runs of gray-levels, reflecting the texture complexity within the image.

- Complexity from the NGTDM category (NGTDM_Comp) indicates the degree of variation within an image. An image is considered complex when it contains numerous primitive components, characterized by non-uniformity and rapid changes in gray-level intensity. A higher Complexity value signifies greater heterogeneity, while a lower value indicates increased uniformity.
- Busyness from the NGTDM category (NGTDM_B) measures the rate of change in intensity between a pixel and its neighbors. A high value indicates a 'busy' image with rapid intensity changes, while lower values correspond to more uniform and homogeneous regions. This feature quantifies the variability of texture within the image, making it particularly useful for characterizing texture and identifying regions with high spatial frequency variations in medical images, such as areas with significant structural detail or noise.
- Coarseness from the NGTDM category (NGTDM_Coar) measures the average difference between the center voxel and its neighborhood, reflecting the spatial rate of change in texture. A higher value indicates a lower rate of spatial change and a more locally uniform texture. However, higher Coarseness values are also associated with greater heterogeneity on a larger scale.
- Contrast from the NGTDM category (NGTDM_Con) measures the spatial intensity changes within an image and is influenced by the overall gray-level dynamic range. High Contrast values indicate a wide dynamic range and significant intensity variations between neighboring voxels, reflecting greater heterogeneity.

**Homogeneous Hyper, Hypo, and Intermediate Intensity Visual Semantic Features.** The following RFs highlight both the homogeneity and intensity of large areas within a lesion, as these substantial regions significantly influence the lesion's overall characteristics. Intermediate feature values typically correspond to PI-RADS 1, indicating a likely benign lesion with uniform and moderate intensity. In contrast, lesions with high-intensity features on DWI and low-intensity features on ADC are associated with PI-RADS 2, reflecting slightly increased suspicion, though still likely benign. The RFs associated with homogeneous hyper-intense, hypo-intense, and intermediate intensity characteristics are outlined in Table 1 (third row) as well as detailed next:

- Large Dependence Low Gray Level Emphasis and Large Dependence High Gray Level Emphasis from the GLDM category (GLDM_LDLGLE and GLDM_LDHGLE) measure the joint distribution of large dependence with lower gray-level values, where large dependence reflects extensive regions of similar pixel intensities, indicating a homogeneous texture. Low gray-level values further indicate uniformity, while in the context of hypo-intensity, GLDM_LDLGLE specifically measures regions with low intensity, often corresponding to darker areas in medical imaging. Conversely, GLDM_LDHGLE emphasizes large zones with high gray-level intensity, with high values indicating homogeneous, hyper-intense lesions. Both features reflect the size and consistency of contiguous regions in the texture.
- Large Area High Gray Level Emphasis and Large Area Low Gray Level Emphasis from the GLSZM category (GLSZM_LAHGLE and GLSZM_LALGLE) measure the emphasis on large zones with specific gray-level intensities within an image. GLSZM_LAHGLE quantifies the proportion of large zones with high gray-level intensity, assigning higher values to extensive, bright, and uniform regions, often associated with well-defined, hyper-intense lesions. Conversely, GLSZM_LALGLE emphasizes large zones with low gray-level intensity, with high values indicating homogeneous, hypo-intense lesions. Both features reflect the size and consistency of contiguous regions in the texture.
- Long Run High Gray Level Emphasis from the GLRLM category (GLRLM_LRHGLE) measures the joint distribution of long run lengths with higher gray-level values. A high value suggests the presence of many long runs of high-intensity pixels in the lesion, indicating texture homogeneity and hyper-intensity. In contrast, the Long Run Low Gray Level Emphasis from the GLRLM category (GLRLM_LRLGLE), with a high value, indicates homogeneity and hypo-intensity.

**Hyper, Hypo, and Intermediate Intensity Visual Semantic Features.** This section highlights RFs that capture intensity and texture variations in ROIs. Features like Maximum intensity (FO_MaxI), The 90th percentile (FO_90P), and The 10th percentile (FO_10P) features from the FO category focus on intensity extremes, aiding in identifying hypo- and hyper-intense regions. Median intensity and Inter Quartile Range features from the FO category (FO_MedI and FO_IQR) emphasize central tendencies, while others provide a broader assessment of intensity distribution, offering comprehensive insights into lesion characteristics. The RFs associated with hyper-intensity, hypo-intensity, and intermediate intensity characteristics are outlined in Table 1 (forth row) as well as detailed below.

- FO_IQR measures variability in the middle 50% of intensity values, where the lower FO_IQR reflects less variability, supporting the homogeneous appearance of PI-RADS 1 lesions.
- The visual semantic feature for PI-RADS 2 is mild hypo-intensity, and these characteristics are captured by specific RFs, providing quantitative descriptions of the lesion. For hypo-intense and hyper-intense lesions, mean intensity from the FO category (FO_MI) measures the average voxel intensity within the lesion, with a lower mean intensity indicating the darker appearance of the lesion, consistent with mild hypo-intensity.
- FO_10P captures the darker portions of the lesion, where a lower value reflects the darkest areas, consistent with mild hypo-intensity.
- FO_90P indicates the brightness of 90% of the intensity values in the region of interest (ROI), where a lower value suggests hypo-intensity.
- Energy and Total Energy from the FO category (FO_E and FO_TE) are directly related to voxel volume from SF category (SF_VV_3D). FO_TE is determined by multiplying SF_VV_3D with FO_E. In this study, since SF_VV_3D is set to {1, 1, 1}, FO_TE is equal to FO_E. Energy measures the magnitude of voxel values in an image, with higher values reflecting a greater sum of squared voxel intensities. ROIs containing hyper-intense voxels (brighter areas) tend to have higher energy values.
- FO_MaxI represents the highest gray-level intensity within the ROI, where the value is lower in mildly hypo-intense lesions compared to surrounding tissues, characterizing the hypo-intense nature.
- FO_MedI reflects the middle gray-level intensity within the ROI, where a lower median indicates that voxel intensities are skewed towards darker values, consistent with mild hypo-intensity.
- Minimum from the FO category (FO_Min) captures the lowest intensity in the image. A higher value indicates a hyper-intense lesion, whereas a lower value does not necessarily indicate a hypo-intense lesion.
- Sum Average from the GLCM category (GLCM_SAv) measures the relationship between pairs of pixels with lower and higher intensity values. It evaluates how frequently dark pixels (low brightness) are paired with bright pixels (high brightness) within the texture. A higher value indicates a more complex and heterogeneous texture, characterized by a mix of dark and bright regions.
- Joint Average from the GLCM category (GLCM_JA) calculates the mean gray-level intensity of the distribution i within the GLCM matrix. It is determined by summing the product of each gray-level intensity and its corresponding joint probability p (i,j). This value reflects the overall brightness or darkness of the image, accounting for the frequency of different gray levels when paired with others.
- Low Gray Level Emphasis from the GLDM category (GLDM_LGLE) measures the prevalence of low gray-levels, reflecting hypo-intense regions with darker appearance due to lower signal intensity. A high value indicates a greater presence of hypo-intense areas.
- High Gray Level Emphasis from the GLDM category (GLDM_HGLE) measures the distribution of higher gray-level values in an image. A higher value indicates a greater concentration of high gray-level intensities, often associated with hyper-intense lesions or regions of elevated signal intensity.
- High Gray Level Zone Emphasis from the GLSZM category (GLSZM_HGLZE) measures the distribution of higher gray-level values. A higher value indicates a greater proportion of zones with high gray-level values, while a lower value suggests hypo-intense or intermediate-intensity lesions.
- Low Gray Level Emphasis from the GLSZM category (GLSZM_LGLZE) measures the distribution of low gray-level size zones within an image. A higher value indicates a greater proportion of low gray-level values and size zones, suggesting increased hypo-intensity in the image.
- High Gray Level Emphasis from the GLSZM category (GLSZM_HGLZE) measures the distribution of high gray-level size zones within an image. A higher value indicates a greater proportion of high gray-level values and size zones, suggesting increased hyper-intensity in the image.
- High Gray Level Run Emphasis from the GLRLM category (GLRLM_HGLRE) measures the distribution of higher gray-level values within runs. A higher GLRLM_HGLRE value indicates a greater concentration of high gray-level intensities, often reflecting the presence of hyper-intense lesions or regions with elevated signal intensity.
- Low Gray Level Run Emphasis from the GLRLM category (GLRLM_LGLRE) measures the distribution of low gray-level values within an image. A higher value indicates a greater concentration of low gray-level values, suggesting that the ROI is predominantly hypo-intense.

**Encapsulated and Circumscribed Visual Semantic Features.** The difference between PI-RADS 1 and PI-RADS 2 lies in their boundaries and contrast patterns. PI-RADS 1 lesions are both circumscribed and fully encapsulated, showing two distinct contrasts: one between the lesion and its fibrous capsule, and another between the capsule and the surrounding tissue, reflecting a very low-risk of malignancy. In contrast, PI-RADS 2 lesions are either incompletely encapsulated or merely circumscribed, with only one contrast between the lesion and the surrounding tissue. Despite often lacking full encapsulation, their well-defined margins still suggest a low probability of cancer.

The RFs associated with encapsulated and circumscribed semantic characteristics are outlined in Table 1 (fifth row) as well as detailed below.
- Run Percentage from the GLRLM category (GLRLM_RP) measures texture coarseness as the ratio of the number of runs to the total number of voxels in the ROI. Lower values indicate a larger proportion of long runs within the ROI, suggesting a circumscribed (well-defined) and uniform texture.
- Strength from the NGTDM category (NGTDM_S) measures the prominence of primitives in an image. A high Strength value indicates that primitives are clearly defined and visible, typically in images with slow intensity changes but large, coarse differences in gray-levels. Such high values suggest the presence of distinct and circumscribed (well-defined) lesions, reflecting significant intensity variations.

**Round, Linear, and Wedge Shape Visual Semantic Features.** SFs describe how closely the ROI resembles a sphere (3D) or a circle (2D), with 3D Compactness 1 (SF_Com1_3D), 3D Compactness 2 (SF_Com2_3D), 3D Spherical Disproportion (SF_SpD_3D), and 2D Spherical Disproportion (SF_SpD_2D) features from the SF category indicating deviations from these ideal shapes. These features are particularly relevant to PI-RADS 1 and occasionally PI-RADS 2. Flatness (SF_F_3D) and Elongation (SF_E_3D) from the 3D SF category, built by Major Axis Length (SF_MAL), Minor Axis Length (SF_MiAL), and Least Axis Length (SF_LAL) features from SF categories, highlight linear and wedge-shaped characteristics, contributing to the geometric understanding of lesions, especially in PI-RADS 2 classifications. The RFs associated with shape characteristics are outlined in Table 1 (sixth row) as well as detailed below.
- SF_F_3D measures the relationship between the largest (major axis) and smallest (least axis) dimensions, where lower values indicate a balanced, round shape.
- SF_E in both 2D and 3D have the same meaning (SF_E_3D and SF_E_2D) represents the relationship between the two largest principal components of the ROI shape. For computational purposes, this feature is defined as the inverse of true elongation. Instead of referring to principal components, it can be described using the lengths of the major and minor axes.
- Surface Area to Volume ratio feature from the 3D SF category (SF_SAVR_3D) provides insight to the compactness of the tumor, with lower ratios suggesting a compact, round shape.
- 3D Sphericity from the SF category (SF_Sp_3D) evaluates how closely the tumor's shape approximates a sphere, while 2D Sphericity (SF_Sp_2D) assesses how closely it resembles a circle. Higher values for these features indicate a round, well-encapsulated appearance.
- SF_Com1_3D quantifies how compact the tumor is compared to a perfect sphere, and SF_Com2_3D measures compactness independent of scale, both supporting the round and regular shape.
- SF_F_3D measures the relationship between the largest and smallest dimensions, where lower values indicate a balanced, round shape.
- SF_SpD_3D describes how the surface area of a tumor compares to the surface area of a sphere with the same volume. It is essentially the inverse of Sphericity. By definition, the value of 3D Spherical Disproportion is always greater than or equal to one, with a value of one representing a perfect sphere. SF_SpD_2D is similar but applies to a 2D context, comparing the shape of a region to a circle instead of a sphere. In T2WI, a score of 1 represents an encapsulated lesion, while a score of 2 represents a circumscribed lesion. Despite their differences, both refer to well-defined structures.
- SF_MAL_2D and 3D represents the longest axis of the lesion, where a longer axis reflects the elongated nature of the wedge-shaped lesion.
- SF_MiAL in both 2D and 3D (SF_MiAL_2D and SF_MiAL_3D) represents the second largest axis, indicating the width or height of the wedge-shaped lesion, contributing to its tapering characteristics.
- SF_LAL_3D represents the smallest axis length, reflecting the thickness of the wedge at its narrowest point, with smaller values indicating a thinner wedge. These axis related features help describe the geometric properties of a wedge-shaped lesion, contributing to a comprehensive understanding of its dimensions and tapering characteristics.

    In the PZ, PI-RADS 1 indicates normal imaging characteristics with no significant intensity variations on DWI and ADC. The visual semantic feature in the PZ is also homogeneous, and the associated RFs are the same as in the TZ. In PZ, a PI-RADS 2 score is associated with a DWI score of 2, where lesions may show linear or wedge-shaped hypo-intensity on ADC and/or linear or wedge-shaped hyper-intensity on high b-value DWI. The prominent visual semantic features include Linear shape and Wedge shape-based features.

### 3.1.2. PI-RADS 3

PI-RADS 3 in the TZ is defined in two scenarios: a T2WI score of 2 requires a DWI score of 4 or 5, whereas a T2WI score of 3 corresponds to a DWI score ranging from 1 to 4. In the first scenario, T2WI scores of 2 in PI-RADS 3 are comparable to T2WI scores of 2 in PI-RADS 2. As previously described, these T2WI scores of 2 are characterized by visual semantic features such as (i) circumscribed (with associated RFs in Table 1, fifth row) and (ii) round, linear, or wedge-shaped (with associated RFs in Table 1, sixth row). In contrast, a DWI score of 4 denotes a focal lesion that is markedly hypo-intense on ADC and markedly hyper-intense on high b-value DWI. The prominent visual semantic features include marked hypo-intensity and hyper-intensity, as previously outlined. A DWI score of 5 shares the characteristics of a DWI score of 4 but also indicates a lesion that measures ≥1.5 cm in its greatest dimension or demonstrates definite extra-prostatic extension or invasive behavior.

For DWI scores of 4 and 5, the most prominent visual semantic features are marked hypo-intensity and marked hyper-intensity. The hypo and hyper-intensity were previously explained in PI-RADS 2 and associated RFs are mentioned in Table 1 (forth row), respectively, with the term 'marked' referring to the higher values of these features. Additionally, the size of the largest dimension plays key role in differentiating a DWI score of 4 from 5, with the largest dimension being directly related to the associated RFs, as outlined in Table 1 (seventh row), as detailed below. Maximum 3D Diameter from the SF category (SF_Max3DD), defined as the largest pairwise Euclidean distance between tumor surface mesh vertices, measuring the largest distance between two points on a tumor's surface in 3D space and directly relating to ≥1.5 cm in greatest dimension criteria. Maximum 2D Diameter from the SF category (SF_Max2DD) in 2D (SF_Max2DD (2D)) and in 3D (SF_Max2DD (Slice, Row, and Column)) is defined as the largest pairwise Euclidean distance between tumor surface mesh vertices within specific anatomical planes. In 3D we have 3 Max2DD the slice plane (i.e. SF_Max2DD (Slice)) row, column, typically axial, represents the largest distance in the axial view. In the column plane (i.e. SF_Max2DD (Column)) row-slice, usually coronal, reflects the largest distance in the coronal view. Similarly, in the row plane (i.e. SF_Max2DD (Row)) column-slice, generally sagittal, it measures the largest distance in the sagittal view. SF_MAL in both 3D and 2D respectively (SF_MAL_2D and SF_MAL_3D), which yields the largest axis length of the ROI-enclosing ellipsoid, is calculated using the largest principal component and indicates the largest dimension of the tumor as represented by the ellipsoid or showing definite extra-prostatic extension or invasive behavior.

In the second scenario, a T2WI score of 3, characterized by heterogeneous signal intensity with obscured margins, corresponds to a DWI score ranging from 1 to 4. Additionally, if a T2WI score cannot be classified as 2, 4, and 5, it is assigned a score of 3. A common characteristic of T2WI score 3 is a heterogeneous signal intensity (heterogeneity) with unclear borders. The definition of heterogeneity is established in PI-RADS score 1, with the corresponding RFs highlighted in Table 1, the second row. Another prominent visual semantic feature of a T2WI score of 3 is an obscured margin, which refers to an unclear boundary. This is the opposite of a circumscribed margin (i.e., well-defined). While obscured and circumscribed margins share similar related RFs, key distinction lies in the feature values: obscured margins have low values, whereas circumscribed margins exhibit high values. A DWI score ranging from 1 to 3 is defined in PI-RADS 2, while a DWI score of 4 is defined in PI-RADS 3 under the first scenario. In PZ, PI-RADS 3 is defined by a DWI score of 3, as explained in PI-RADS 2 (TZ) and DCE negative which is assessed by the radiologist. For DWI score 3, the prominent visual semantic features focal hypo-intensity on ADC and/or focal hyper-intensity on high b-value DWI, with the possibility of being markedly hypo-intense on ADC or markedly hyper-intense on high b-value DWI, but not applicable to both. DCE positive is a binary task assessed by the radiologist.

### 3.1.3. PI-RADS 4 & 5

PI-RADS 4 in the TZ is defined in two scenarios: a T2WI score of 3 combined with a DWI score of 5 or a T2WI score of 4. The T2WI score of 3 and DWI score of 5 have been defined previously, along with their visual semantic features and related RFs. In the first scenario, T2WI scores of 3 in PI-RADS 4 are comparable to T2WI scores of 3 in PI-RADS 3. As previously described, these T2WI scores of 3 are characterized by visual semantic features such as: (i) heterogeneity (with associated RFs in Table 1, second row) and (ii) circumscribed (with associated RFs in Table 1, fifth row). The obscured margin is the opposite of a circumscribed margin (i.e., well-defined). Moreover, a DWI score of 5 denotes a focal lesion characterized by marked hypo-intensity and hyper-intensity (outlined in Table 1, forth row), with the additional criteria of measuring ≥1.5 cm in its greatest dimension (as outlined in seventh row) or exhibiting definite extra-prostatic extension or invasive behavior. In the second scenario, a T2WI score of 4 is characterized by a lenticular (Table 1, sixth row), non-circumscribed (or obscured margins, fifth row), moderately homogeneous Hypo-intense (third row) lesions that are less than 1.5 cm in its greatest dimension. The prominent visual semantic features include lenticular shape, which refers to having the shape of a bi-convex lens or crescentic. The related RFs for lenticular shape include both SF_Sp_2D and SF_Sp_3D features, which can make the lesion appear more spherical depending on the angle of observation, SF_F_3D, as lenticular shapes typically have a certain degree of flattening compared to a perfect sphere, and SF_E_3D, where if viewed from an edge-on perspective, the lesion may resemble a straight line due to its flattened nature, a common characteristic of ellipsoidal or lenticular bodies.

In the PZ, PI-RADS 4 is defined by two scenarios. In the first scenario, a DWI score of 3, as described in PI-RADS 2 (TZ), combined with a positive DCE result assessed by the radiologist. For a DWI score of 3, the prominent visual semantic features include focal hypo-intensity on ADC and/or focal hyper-intensity on high b-value DWI. While the lesion may be markedly hypo-intense on ADC or markedly hyper-intense on high b-value DWI, it does not exhibit both characteristics simultaneously. A positive DCE is a binary assessment performed by the radiologist. In the second scenario, PI-RADS 4 corresponds to a DWI score of 4, as elaborated in PI-RADS 3 under the first scenario. PI-RADS 5 in TZ is associated with a T2WI score of 5, which is defined similarly to T2WI score of 4 but with the lesion being ≥1.5 cm in greatest dimension or showing definite extra-prostatic extension or invasive behavior. In PZ, PI-RADS 5 is defined by a DWI score of 5, which is the same as DWI score 4 but with the lesion being ≥1.5 cm in greatest dimension or showing definite extra-prostatic extension or invasive behavior. No meaningful relationship was found between the RFs from the SF category, such as Mesh Surface (SF_MS_2D), Pixel Surface (SF_PS_2D), Perimeter (SF_P_2D), and Perimeter to Surface ratio (SF_PSR_2D), and visual semantic features.

**Table. 1.** Radiomics features related to PI-RADS scores of 1 to 5

| Visual Semantic Feature | Related Radiomics Feature |
|---|---|
| **Heterogeneity and Uniformity** | FO_En, FO_SD, FO_V, FO_MAD, FO_RMAD, FO_RMS, FO_Ku, FO_R, GLCM_IDM, GLCM_IDMN, GLCM_JEn, FO_Un, FO_Sk, GLCM_ID, GLCM_IDN, GLCM_JE, GLCM_CT, GLCM_Corr, GLCM_Co, GLCM_CP, GLCM_CS, GLCM_AC, GLCM_DiEn, GLCM_IMC1 , GLCM_IMC2, GLCM_MCC, GLCM_DA, GLCM_SEn, GLCM_SQ, GLCM_MP, GLCM_IV, GLDM_DN, GLDM_DNN, GLDM_DV, GLDM_GLN, GLDM_DeEn, GLDM_GLV, GLDM_DV, GLDM_LDE, GLDM_SDLGLE, GLDM_SDHGLE, GLDM_SDE, GLSZM_GLN, GLSZM_GLNN, GLSZM_SZN, GLSZM_SZNN, GLSZM_ZP, GLSZM_SAE, GLSZM_SALGLE, GLSZM_SAHGLE, GLSZM_ZE, GLSZM_GLV, GLSZM_LAE, GLSZM_ZV, GLRLM_SRHGLE, GLRLM_SRE, GLRLM_LRE, GLRLM_RLN, GLRLM_RLNN, GLRLM_SRLGLE, GLRLM_RE, GLRLM_GLN, GLRLM_GLNN, GLRLM_GLV, GLRLM_RV, NGTDM_Comp, NGTDM_B, NGTDM_Coar, NGTDM_Con |
| **Homogeneous Hyper, Hypo, and Intermediate Intensity** | GLDM_LDLGLE, GLSZM_LAHGLE, GLSZM_LALGLE, GLRLM_LRHGLE, GLRLM_LRLGLE |
| **Hyper, Hypo, and Intermediate Intensity** | FO_IQR, FO_MI, FO_10P, FO_90P, FO_E, FO_TE, SF_VV, FO_MaxI, FO_MedI, FO_Min, GLCM_Sav, GLCM_Sav, GLCM_JA, GLDM_LGLE, GLDM_HGLE, GLSZM_HGLZE, GLSZM_LGLZE, GLSZM_HGLZE, GLRLM_HGLRE, GLRLM_LGLRE |
| **Encapsulated and Circumscribed** | GLRLM_RP, NGTDM_Str |
| **Round, Linear, Wedge, and Lenticular Shape** | SF_F_3D, SF_E_3D, SF_SAVR_3D, SF_Sp_3D, SF_Sp_2D, SF_Com1_3D, SF_Com2_3D, SF_F_3D, SF_SpD_3D, SF_SpD_2D, SF_MAL_2D, SF_MAL_3D, SF_MiAL_2D, SF_MiAL_3D, SF_LAL |
| **Largest Dimension** | SF_Max3DD, SF_Max2DD such as SF_Max2DD (Slice), SF_Max2DD (Column), SF_Max2DD (Row) |

### 3.2. Interpretable and Explainable Classification Task for UAI

As shown in Figure 3, combining T2WI, DWI, and ADC imaging sequences with three feature selection algorithms—AFT, CCF and FSF feature selection algorithms—and utilizing LRC achieved the highest average accuracy of 0.78 ± 0.01, with an external nested testing accuracy of 0.74 ± 0.02. As shown in Table 2, all these FSAs selected similar RFs such as T2WI_FO_90P (FO_90P extracted from T2WI sequence), T2WI_FO_V (FO_V extracted from T2WI sequence), ADC_SF_LAL_3D (SF_LAL_3D extracted from ADC sequence), ADC_SF_SAVR_3D (SF_SAVR_3D extracted from ADC sequence), and ADC_GLRLM_RE (GLRLM_RE extracted from ADC sequence). Moreover, some other combinations of these three feature selections linked with CBC, LDA, and STC reached a close performance of 0.76. In addition, when using T2WI alone, some ML algorithms achieved a maximum accuracy of 0.73 in five-fold cross-validation. Specifically, the combination of AFT and NBC reached an accuracy of 0.73 ± 0.06, with external testing yielding 0.71 ± 0.01, utilizing the selected features SF_MAL_2D, SF_Max2DD (Column), SF_Max2DD (Slice), SF_Max3DD, and GLDM_LDLGLE. Mixture of CST and MLP with selected features of surface area from the SF category (SF_SA), GLDM_LDLGLE, GLRM_GLN, GLSZM_LAE, and GLSZM_LAHGLE arrived at average five-fold cross-validation and external tests of 0.73. In addition, combining both the CCF and FSF with MLP and NBC using the similar features selected by CST achieved an average accuracy of 0.73 and an external testing accuracy exceeding 0.71. Finally, the combination of LAS with LDA, using the selected features such as SF_Max2DD (Column), SF_SAVR_3D, FO_Ku, GLDM_LDLGLE, and NGTDM_S, reached an average accuracy of 0.73 and an external testing accuracy of 0.71.

Moreover, when using DWI alone, the combination of AFT with LRC and MLP achieved an average accuracy of $0.72 \pm 0.06$ and an external testing accuracy exceeding 0.72, utilizing the selected features such as SF_LAL_3D, SF_MiAL_3D, GLDM_GLN, GLRM_GLN, and GLRLM_REn. Additionally, combining CST with LDA, LRC, MLP, NBC, and SVM achieved an average accuracy of 0.72 and an external testing accuracy exceeding 0.70, using the selected features such as SF_MV_3D, SF_VV_3D, GLSZM_LAE, GLSZM_LAHGLE, and GLRLM_LALGLE. CCF and FSF feature selection algorithms in Combining with LRC and MLP resulted in an average accuracy of 0.72 and an external testing accuracy exceeding 0.72, utilizing the similar features by AFT. IGS linked with LRC and MLP achieved an average accuracy of 0.72 and an external testing accuracy exceeding 0.73, using the selected features such as SF_Max2DD (Slice), SF_SA_3D, FO_TE, GLCM_MP, and NGTDM_Con Additionally, MIS and UFS combined with NBC and SVM yielded an average accuracy of 0.72 and an external testing accuracy exceeding 0.69, using the selected features such as SF_SA_3D, FO_TE, GLCM_JE, and GLCM_MP, GLRM_SRLGLE. Finally, when using ADC alone, IGS and MIS combined with CBC, utilizing the selected features such as SF_Max3DD, SF_MV_3D, SF_SAVR_3D, GLSZM_LAE, GLSZM_LGLZE, achieved an average accuracy of 0.73 and an external testing accuracy exceeding 0.68. Furthermore, combining UFS with XGB, RFC, and STC using selected features such as SF_Max3DD, SF_MV_3D, SF_SAVR_3D, GLRM_SRLGLE, and GLSZM_LAE yielded a performance of 0.73 and an external testing accuracy exceeding 0.65. There is a significant difference between the best performance of 0.78, achieved by combining three imaging sequences with FSAs such as AFT, CCF, and FSF, and utilizing LRC classifier, compared to the high performances obtained from individual datasets (Paired-t test, p-value <0.05). Moreover, in this study, subsets of selected features comprising 5, 10, 20, 40, 80, and 100 features were examined using FSAs. The results indicated that the highest accuracy of 0.78 was achieved with just five relevant features, and no further improvement was observed as the number of features increased.

**Table 2.** Selected features by different feature selection algorithms (FSA) applied on different datasets including T2-weighted imaging (T2WI) alone, diffusion-weighted imaging (DWI) alone, and apparent diffusion coefficient (ADC) alone and a mixture of T2WI, DWI, ADC. CST: Chi-Square Test, CCF: Correlation Coefficient, MIS: Mutual Information, VTS: Variance Threshold, AFT: ANOVA F-test, IGS: Information Gain, UFS: Univariate Feature Selection, FSF: Fisher Score, LAS: Least Absolute Shrinkage and Selection Operator.

| FSA | T2WI | DWI | ADC | T2WI+DWI+ADC Sequence name + Feature name |
|---|---|---|---|---|
| CST | SF_SA_3D, GLDM_LDLGLE, GLRM_GLN, GLSZM_LAE, GLSZM_LAHGLE | SF_MV_3D, SF_VV_3D, GLSZM_LAE, GLSZM_LAHGLE, GLSZM_LALGLE | GLCM_AC, GLCM_JA, GLCM_SAv, GLDM_HGLE, GLRLM_HGLRE | ADC_GLCM_AC, ADC_GLCM_JA, ADC_GLCM_SAv, ADC_GLDM_HGLE, ADC_GLRLM_HGLRE |
| CCF | GLDM_LDLGLE, SF_MAL_2D, SF_Max3DD, SF_Max2DD (Slice), SF_Max2DD (Column) | GLDM_GLN, GLRM_GLN, GLRLM_REn, SF_MiAL_3D, SF_LAL_3D | GLRLM_RLN, GLDM_LDHGLE, SF_SAVR_3D, GLRLM_REn, SF_LAL_3D | T2WI_FO_V, T2WI_FO_90P, ADC_SF_SAVR_3D, ADC_GLRLM_REn, ADC_SF_LAL_3D |
| MIS | GLDM_DE, GLDM_LDE, GLRLM_RLNN, GLRLM_RP, GLSZM_LAE | SF_SA_3D, FO_TE, GLCM_JE, GLCM_MP, GLRIM_SRLGLE | SF_Max3DD, SF_MV_3D, SF_SAVR_3D, GLSZM_LAE, GLSZM_LGLZE | DWI_SF_Max3DD, DWI_GLCM_CS, ADC_SF_Max3DD, ADC_SF_SAVR_3D, ADC_GLSZM_LAE |
| VTS | GLCM_JEn, GLCM_JE, FO_Un, GLDM_GLV, GLCM_SQ | GLDM_LGLE, GLDM_HGLE, GLCM_JA, GLCM_SAv, NGTDM_Coar | GLDM_LGLE, GLDM_HGLE, GLCM_SAv, GLCM_JA, NGTDM_Coar | ADC_GLDM_LGLE, ADC_GLCM_SAv, ADC_GLCM_JA, DWI_NGTDM_Coar, ADC_NGTDM_Coar |
| AFT | SF_MAL_2D, SF_Max2DD (Column), SF_Max2DD (Slice), SF_Max3DD, GLDM_LDLGLE | SF_LAL_3D, SF_MiAL_3D, GLDM_GLN, GLRM_GLN, GLRLM_REn | SF_LAL_3D, SF_SAVR_3D, GLDM_LDHGLE, GLRLM_RE, GLRLM_RLN | T2WI_FO_90P, T2WI_FO_V, ADC_SF_LAL_3D, ADC_SF_SAVR_3D, ADC_GLRLM_REn |
| IGS | GLDM_DE, GLDM_LDE, GLRLM_RLNN, GLRLM_RP, GLSZM_LAE | SF_Max2DD (Slice), SF_SA_3D, FO_TE, GLCM_MP, NGTDM_Con | SF_Max3DD, SF_MV_3D, SF_SAVR_3D, GLSZM_LAE, GLSZM_LGLZE | DWI_SF_F_3D, DWI_GLCM_CS, ADC_SF_Max3DD, ADC_SF_SAVR_3D, ADC_GLSZM_LAE |
| UFS | GLDM_DE, GLDM_LDE, GLRLM_RLNN, GLRLM_RP, GLSZM_LAE | SF_SA, FO_TE, GLCM_JE, GLCM_MP, GLRIM_SRLGLE | SF_Max3DD, SF_MV_3D, SF_SAVR_3D, GLRIM_SRLGLE, GLSZM_LAE | DWI_SF_F_3D, DWI_SF_Max2DD (Row), ADC_SF_Max3DD, ADC_SF_SAVR_3D, ADC_GLSZM_LAE |
| FSF | GLDM_LDLGLE, SF_MAL_2D, SF_Max3DD SF_Max2DD (Slice), SF_Max2DD (Column) | GLDM_GLN, GLRM_GLN, GLRLM_REn, SF_MiAL_3D, SF_LAL_3D | GLRLM_RLN, GLDM_LDHGLE, SF_SAVR_3D, GLRLM_REn, SF_LAL_3D | T2WI_FO_V, T2WI_FO_90P, ADC_SF_SAVR_3D, ADC_GLRLM_REn, ADC_SF_LAL_3D |
| LAS | SF_Max2DD (Column), SF_SAVR_3D, FO_Ku, GLDM_LDLGLE, NGTDM_S | SF_LAL_3D, SF_MiAL_3D, GLCM_CS, GLRLM_REn, NGTDM_S | GLDM_HGLE, GLDM_LDE, GLDM_LGLE, GLRLM_REn, NGTDM_S | ADC_GLRLM_REn, T2WI_GLCM_JA ADC_GLDM_HGLE, T2WI_GLSZM_SAHGLE, T2WI_GLCM_SQ |

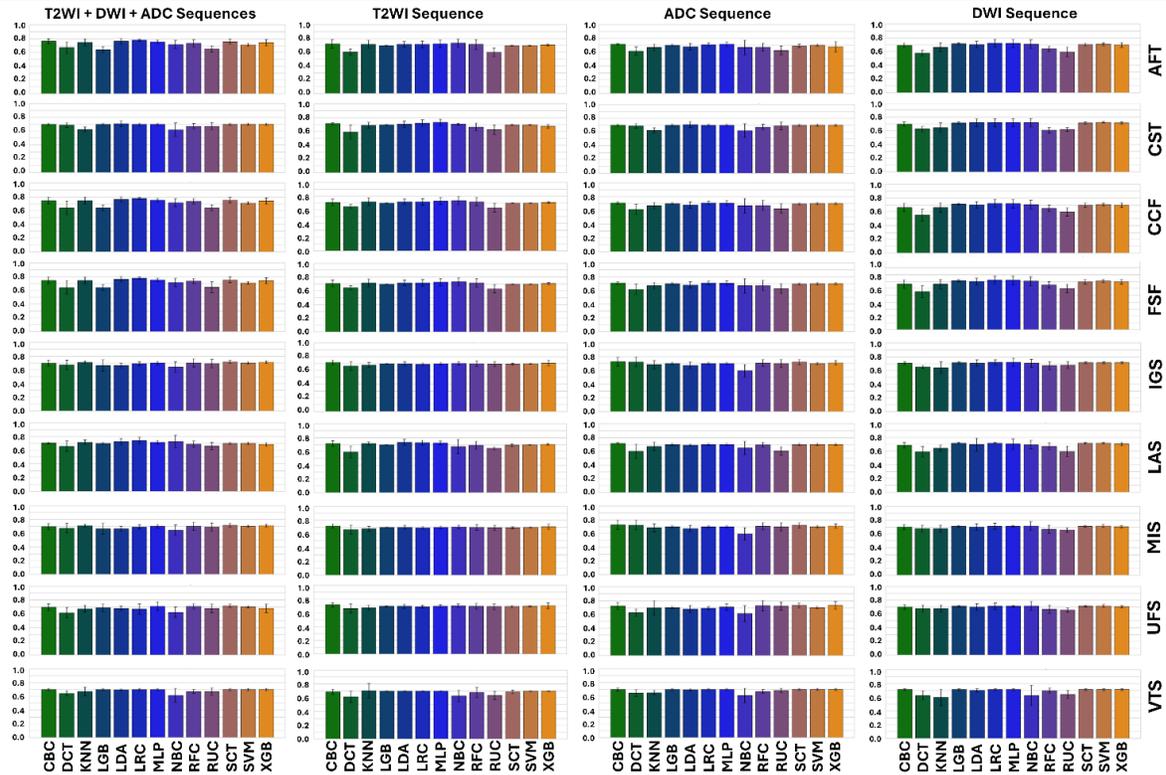

**Fig. 3.** Bar plot of Mean ± standard deviation of feature selection algorithms applied to classification algorithms (mentioned in 1.2). T2-weighted imaging (T2WI) alone, diffusion-weighted imaging (DWI) alone, and apparent diffusion coefficient (ADC) alone and a mixture of T2WI, DWI, ADC. CST: Chi-Square Test, CCF: Correlation Coefficient, MIS: Mutual Information, VTS: Variance Threshold, AFT: ANOVA F-test, IGS: Information Gain, UFS: Univariate Feature Selection, FSF: Fisher Score, LAS: Least Absolute Shrinkage and Selection Operator, DTC: Decision Tree Classification, LRC: Logistic Regression, LDA: Linear Discriminant Analysis, NBC: Naive Bayes Classifier, KNN: K-Nearest Neighbors, RUC: RuleFit Classifier, RFC: Random Forest, XGB: XGBoost Classifier, LGB: LightGBM Classifier, CBC: CatBoost Classifier, SVM: Support Vector Machine, STC: Stacking Classifier, and MLP: Multilayer Perceptron.

## 4. DISCUSSION

Adoption of AI in clinical practice can be hindered by challenges in interpretability and explainability, particularly due to the "black box" nature of AI models. PI-RADS standardizes multiparametric prostate MRI interpretation, categorizing lesions from 1 to 5 to assess cancer likelihood. Integrating RFs with PI-RADS scoring enhances diagnostic precision by providing quantitative metrics that support radiological assessments. By developing a standardized RFs dictionary aligned with PI-RADS and incorporating interpretable ML algorithms, we improve both the interpretability and explainability of AI models in prostate cancer diagnostics. This approach not only boosts diagnostic accuracy but also builds radiologist's trust, making AI tools more viable for routine medical use across various health fields.

PI-RADS 1 and 2 typically represent normal or benign findings in both the PZ and TZ of the prostate. PI-RADS 1 lesions are characterized by homogeneous signal intensity and well-defined, round shape on T2WI. RFs such as high homogeneity measures and shape-based features indicating sphericity and compactness quantitatively confirm the benign nature of these lesions. For instance, features that assess local homogeneity and uniformity capture the consistent signal intensity, while shape RFs measure how closely the lesion's form approximates a sphere. PI-RADS 2 lesions may appear mildly hypo-intense on T2WI and could be presented as encapsulated or atypical nodules. They might also display linear or wedge-shapes, particularly in the PZ. Texture-based RFs that indicate mild intensity variations capture these characteristics, suggesting a low likelihood of clinically significant cancer. SFs reflecting linear or wedge-forms quantitatively describe the geometric properties of these lesions, providing further insight into their benign nature.

PI-RADS 3 represents an equivocal category where the likelihood of clinically significant cancer is uncertain. Lesions may exhibit mildly hypo-intense signals and adopt more complex shapes like linear or wedge-forms on T2WI. Texture-based RFs capturing heterogeneity—such as measures of entropy and intensity variation—along with Shape-Features indicating elongation or irregularity, help quantify these ambiguous imaging features. The integration of these RFs aids in differentiating PI-RADS 3 lesions from both lower and higher grade lesions by providing objective metrics that reflect their uncertain nature. In the TZ, a PI-RADS 3 lesion could be defined by a T2WI score of 2 with a DWI score of 4 or 5, or a T2WI score of 3 with a DWI score ranging from 1 to 4. The heterogeneous signal intensity is a prominent feature here, captured by RFs that measure entropy and intensity variations. In the PZ, a PI-RADS 3 lesion is typically associated with a DWI score of 3 and negative DCE imaging. Texture RFs related to hypo-intensity and hyper-intensity on DWI and ADC maps provide quantitative support for these assessments. Higher scores, PI-RADS 4 and 5, indicate a greater probability of malignancy. In the TZ, a PI-RADS 4 lesion is characterized by a lenticular or non-circumscribed shape, homogeneous moderately hypo-intense signal on T2WI, and measures less than 1.5 cm in greatest dimension. Shape-based RFs quantify the lenticular form and size, while texture RFs confirm the homogeneous signal intensity.

A PI-RADS 5 lesion in the TZ is similar but either equals or exceeds 1.5 cm in size or shows definite extra-prostatic extension or invasive behavior. Shape-based RFs like SF_Max3DD and SF-MAL-2D and ED quantify lesion size, directly relating to the 1.5 cm threshold that distinguishes PI-RADS 4 & 5 lesions. In the PZ, a PI-RADS 4 lesion is defined by a DWI score of 4, indicating a focal lesion that is markedly hypo-intense on the ADC map and markedly hyper-intense on high b-value DWI. Texture RFs measuring intensity extremes and heterogeneity support these imaging findings. A PI-RADS 5 lesion in the PZ is similar to a PI-RADS 4 lesion but either measures equal to or greater than 1.5 cm or exhibits definite extra-prostatic extension or invasive behavior. Again, shape-based RFs quantifying size and irregularity provide objective metrics that support the PI-RADS classification. Integrating RFs with PI-RADS scoring enhances diagnostic accuracy by providing quantitative data that complement qualitative assessments. Texture-based RFs derived from statistical analyses—such as GLCM and GLSZM—measure aspects like homogeneity, entropy, and intensity variation within lesions. These features help differentiate between uniform and heterogeneous tissue characteristics, which are crucial in identifying malignant tissues. For example, a high entropy value might indicate a heterogeneous lesion, raising suspicion for malignancy, while high homogeneity measures suggest benign tissue.

Shape-based RFs assess geometric properties, offering insights into the lesion's form and potential aggressiveness. Irregular or elongated shapes captured by specific RFs may indicate malignant behavior, while well-defined, spherical shapes are more suggestive of benign conditions. Size-related features like SF_Max3DD and SF_MAL_2D and 3D are particularly important in distinguishing between PI-RADS 4 and 5 lesions, as the 1.5 cm threshold is a critical factor in the PI-RADS classification. The use of RFs also holds promise for personalized medicine. By providing a comprehensive quantitative assessment of lesion characteristics, RFs can help in risk stratification and treatment planning. For instance, patients with lesions that have high-risk radiomic profiles might benefit from more aggressive treatment plans, while those with low-risk profiles could be candidates for active surveillance [101, 102].

Therefore, the integration of RFs with the PI-RADS scoring system enhances the evaluation of prostate lesions by providing objective, quantifiable data that support and refine radiological/biological assessments. This approach has the potential to improve diagnostic accuracy, assist in risk stratification, and ultimately lead to more personalized and effective patient management strategies. Moreover, identifying these relationships between RFs and PI-RADS scoring enhances the interpretability and explainability of AI algorithms used in prostate cancer imaging. By correlating specific quantitative features with established diagnostic categories, AI models can provide transparent reasoning behind their predictions, making it easier for radiologists to understand and trust the technology. This interpretability is crucial in a clinical setting, as it allows healthcare professionals to validate AI-generated assessments against known medical criteria, ensuring that decisions are based on relevant and understandable factors. Ultimately, this synergy between RFs and PI-RADS not only improves diagnostic accuracy but also can foster greater confidence in AI-assisted tools, leading to more personalized and effective patient treatment.

To address the challenges of UAI, this study presents examples utilizing our radiological and biological dictionary of RFs in conjunction with IAI algorithms. In our study, combining RFs extracted from T2WI, DWI, and ADC sequences with FSAs like AFT, CCF, and FSF, and utilizing LRC, resulted in the highest average accuracy of 0.78. This performance was significantly better than using individual imaging sequences alone (paired t-test, $p < 0.05$). Notably, the optimal accuracy was achieved with just five selected features, and no further improvement was observed with an increased number of features. This suggests that a small, well-selected subset of features can be highly effective for classification tasks in medical imaging. The results highlight the importance of integrating multiple imaging modalities and employing robust feature selection algorithms to enhance the predictive accuracy of ML models. The best selected feature subsets including T2WI_FO_90P and T2WI_FO_V extracted from T2WI sequence as well as SF_LAL_3D, SF_SAVR_3D, and GLRLM_RE extracted from ADC.

T2WI_FO_90P represents the brightness level below which 90% of intensity values in the ROI fall; lower values suggest hypo-intensity cause of T2WI selection is due to better visualization of signal intensity in this sequence. In PI-RADS scoring, T2WI is crucial for evaluating lesions in the TZ, where heterogeneity indicate higher scores and increased prostate cancer risk. T2WI_FO_V, measures the spread of intensity values within a lesion, making it particularly useful in T2WI for assessing tissue heterogeneity. High variance, along with other RFs in T2WI, can suggest areas of abnormal intensity variations (heterogeneity). For instance, TZ heterogeneity has been linked to age-related changes such as benign prostatic hyperplasia or prostate cancer. These variations in tissue characteristics help in understanding different conditions within the prostate, from benign growth to potential malignancies. ADC_SF_LAL-3D represents flatness, which is calculated as the square root of the ratio of the smallest principal component (λ least) to the largest principal component (λ major). This ratio serves as an indicator to assess lenticular shape and clinical wedge shape. In the case of a spherical shape (the lowest-risk lesion shape), the lengths of the least, major, and minor axes are equal. Along with other features, these characteristics are valuable in clinical assessments for stratifying patients into high and low-risk categories. The feature ADC_SF_SAVR_3D varies based on the shape and compactness of a lesion. For example, a spherical shape has the highest compactness and lowest risk, but as it flattens, it becomes lenticular, and with further flattening, the ADC_SF_SAVR_3D value increases, indicating a more linear shape. In our study, linear shapes are associated with low-risk, while lenticular shapes are linked to higher-risk for specific lesions. ADC_GLRLM_REn measures the variation in run lengths within the texture of a lesion, capturing irregular patterns and consistency in gray-levels. This feature represents the heterogeneity and homogeneity of lesions, especially in ADC that represent diffusion of water molecules within tissue; it can be a good indication of tissue structure specifically heterogeneity and overall helps in understanding different conditions within the prostate, from benign growth to potential malignancies [103].

Using T2WI alone, some sets with the following RFs achieved a performance score of 0.73. SF_MAL-3D, SF_Max2DD (Column), SF_Max2DD (Slice), and SF_Max3DD represent the maximum diameter of the lesion in three-dimensional space; together, they provide valuable information about tumor direction and spread, helping differentiate very high-risk lesions from high-risk ones and assess the risk of extra-prostatic invasion. SF_SAVR_3D is also applicable here. In T2WI, it is more precise and valuable because T2WI provides better prostate anatomical details, effectively aiding to understand lesion's anatomy. SF_SA_3D represents the surface area of a lesion, offering critical insights into its boundary and surface characteristics. Larger surface area measurements often correlate with more irregular and potentially malignant lesions, aiding in the distinction between well-defined benign structures and irregular malignant ones. When analyzed in conjunction with features like SF_SAVR_3D and volume, the surface area provides a more comprehensive understanding of the lesion's morphology, enabling a deeper and more reliable assessment compared to evaluating surface area in isolation. FO_Ku, measures the "peakedness" of intensity distributions within the lesion, highlighting tissue texture homogeneity or heterogeneity. T2WI, with its excellent ability to distinguish subtle intensity variations, is particularly effective for visualizing kurtosis. Lower values indicate more uniform structures (e.g., benign nodules), whereas high kurtosis suggests heterogeneity—research shows that heterogeneous prostates on MRI are associated with more aggressive pathology [104].

GLDM_LDLGLE highlights the distribution of low gray-level values across dependent regions, helping identify large, homogeneous hypo-intense areas; this often relates to low-risk lesions matching T2WI score 2, which describes homogeneous, mildly hypo-intense areas. GLRM_GLN measures the uniformity of consecutive gray-levels in runs, capturing textural uniformity within a lesion; T2WI's contrast enhanced imaging enable clear visualization of this feature, which is beneficial for identifying high-risk lesions with irregular, non-uniform texture patterns in the PZ or TZ. Lower values indicate consistent textures, often seen in benign tissues, while higher values suggest textural heterogeneity. GLSZM_LAE identifies larger, contiguous zones of similar intensity within the lesion; T2WI's high-resolution depiction of tissue structure is ideal for capturing this feature. Higher values indicate larger, uniform zones often associated with benign lesions, whereas lower values, reflecting smaller, more variable zones, may indicate malignant heterogeneity. GLSZM_LAHGLE captures large zones of high intensity values within a lesion, indicative of homogeneous but hyper-intense areas; T2WI's contrast sensitivity allows detection of these zones, particularly in high-grade tumors exhibiting uniform but intense textures. Strength from NGTDM category (NGTDM_S) quantifies the intensity difference between a voxel and its neighbors, capturing overall lesion texture strength; T2WI's enhanced contrast is particularly useful for observing these neighborhood relationships and detecting strong texture contrasts in high-risk lesions. High values are often associated with irregular, heterogeneous textures common in malignant lesions.

Using ADC alone, certain ML algorithms achieved an average accuracy of 0.73 by utilizing selected features such as those listed below. SF_Max3DD, SF_SAVR_3D, and GLSZM_LAE have been elaborated above. Additionally, SF_MV-3D measure the volume of ROI. SF-VV-3D is calculated from the triangle mesh of the ROI. SF_MV-3D quantifies the size of a prostate tumor, which is essential for assessing cancer aggressiveness and guiding treatment. Larger volumes are associated with higher risks of recurrence and metastasis, aiding to risk stratification and treatment planning [105]. GLSZM_LGLZE measures the distribution of lower gray-level size zones, with higher values indicating a greater proportion of low gray-level zones, which may correlate with hypo-intense, suspicious lesions on ADC. Similarly, GLRM_SRLGLE assesses the distribution of shorter run lengths with lower gray-level values, where higher values suggest increased heterogeneity. Together, these metrics provide insights into lesion characteristics, with GLSZM_LGLZE highlighting regions of hypo-intensity and GLRM_SRLGLE capturing texture irregularities, both of which are valuable for identifying and characterizing suspicious lesions.

With DWI alone, specific ML algorithms reached an average accuracy of 0.72 using the selected features outlined below. SF_LAL-3D, GLRM_GLN, GLRM_SRLGLE, SF_Max2DD (Slice), SF_SA-3D, SF_MV-3D, GLRLM_REn, GLSZM_LAE, and GLSZM_LAHGLE have been detailed above. Furthermore, SF_MiAL_3D represents the second largest axis length of the ROI-enclosing ellipsoid, indicating elongation; higher values suggest a linear shape, often linked to lower-risk lesions. GLRLM_LALGLE highlights the distribution of low gray- level values across regions, identifying large, homogeneous hypo-intense areas. This feature is associated with low-risk, mildly hypo-intense regions, consistent with T2WI score 2. NGTDM_Con measures spatial intensity changes in an image by assessing gray-level variations between neighboring voxels. Higher contrast values indicate greater heterogeneity and significant differences between adjacent regions, correlating with a higher-risk of malignancy. Lower values suggest a more uniform gray-level distribution across the image. FO_TE is calculated by scaling the Total Energy by the voxel volume (SF_VV-3D), making it a good indicator of lesion intensity. GLDM_GLN measures the similarity of gray-level intensity values in an image. A lower value correlates with greater similarity in intensity values, indicating higher homogeneity. Therefore, a lower value represents a more benign lesion. GLCM_MP measures the frequency of the most common neighboring intensity pair in an image. Higher values indicate greater homogeneity, correlating with a lower-risk lesion. GLCM_JE quantifies homogeneous patterns in an image. Higher values indicate more frequent occurrences of certain gray-level pairs, suggesting a more uniform texture and a lower-risk lesion.

As evidenced in our study, collaboration between AI developers, medical physicists and healthcare professionals is essential for creating understandable AI solutions in medical imaging. Such partnerships enable the development of a biological and radiological RFs dictionary that links imaging features to clinical interpretations, enhancing model transparency and trustworthiness. By aligning AI outputs with medical standards, these solutions become more reliable and suitable for clinical adoption, supporting informed decision making and fostering clinician confidence. This framework should include creating RFs dictionaries for other cancers, such as breast cancer, lung cancer, lymphoma and other, as well as being adopted for additional modalities like PET, CT, mammography, and SPECT. Tailoring these dictionaries to each disease and modality would improve diagnostic accuracy and relevance. Since Pyradiomics supports around 119 RFs, whereas ViSERA software supports over 400 RFs, we aim to incorporate the additional features extracted by ViSERA into our feature dictionary [106]. The limitation of this study is the dataset size, which may affect the generalizability of the obtained results. A multicenter study with data from different scanners could improve model robustness, as variations in imaging protocols can impact feature consistency. Additionally, expanding the range of IAI/XAI methods, along with complex models, would further enhance validation. The RFs dictionary and relationship findings would also benefit from expert validation to ensure clinical applicability and reliability. Involving a diverse panel of radiologists, oncologists, and imaging specialists could refine the dictionary by incorporating varied clinical perspectives, establishing standardized interpretations, and improving consistency across institutions. This comprehensive validation would increase the model's diagnostic accuracy and reliability, leading to clinical excellence.

## 5. CONCLUSION

This work addresses UAI challenges in medical imaging by integrating a radiological/biological RFs dictionary with IAI algorithms. The RFs dictionary aligns AI outputs with diagnostic standards like PI-RADS, improving interpretability for clinicians. However, IAI/XAI still requires intervention from clinical experts to interpret results and lacks standardized guidelines or radiological/biological RFs dictionaries. Bridging these gaps is essential to ensure both technical and practical understanding for experts and stakeholders. Key features from T2WI and ADC MRI imaging sequences—T2WI_FO_90P, capturing hypo-intensity relevant to prostate cancer risk; T2WI_FO_V, indicating lesion heterogeneity; ADC_SF_LAL_3D and ADC_SF_SAVR_3D, describing lesion shape and compactness; and GLRLM_REn, reflecting texture consistency—achieved a high accuracy of 0.78 for prediction of the UCLA outcomes in prostate cancer patients, significantly outperforming single-sequence approaches (p-value< 0.05). AI developers can use this RFs dictionary to build more UAI solutions, while clinical professionals can rely on it for easier interpretation of AI model outputs in assessments. Serving as a common language, this dictionary can foster collaboration between clinical professionals and AI developers, advancing the creation of robust, trustworthy AI solutions that support reliable, interpretable clinical decisions.

**DATA AND CODE AVILABILITY**. All codes and tables are publicly shared at:
https://github.com/MohammadRSalmanpour/Biological-and-Radiological-Dictionary-of-Radiomic-Features/tree/main

**ACKNOWLEDEGEMENTS.** This work was supported by the Mitacs Accelerate program grant number AWD-024298-IT33280. We acknowledge the support of the Natural Sciences and Engineering Research Council of Canada (NSERC). Nous remercions le Conseil de recherches en sciences naturelles et en génie du Canada (CRSNG) de son soutien. Discovery Grant RGPIN-2023-03575. We also acknowledge the support from Microsoft's AI for Good Lab. This study was also supported by the Natural Sciences and Engineering Research Council of Canada (NSERC) Discovery Grant RGPIN-2019-06467, as well as the Technological Virtual Collaboration Corporation (TECVICO CORP.), Vancouver, Canada.

**CONFILICT OF INTREST.** The authors have no relevant conflicts of interest to disclose.

**REFERENCES**

[1]  Z. C. Lipton, "The Mythos of Model Interpretability," *ACM,* 2018.

[2]  C. Rudin, "Stop explaining black box machine learning models for high stakes decisions and use interpretable models instead," *Nature Machine Intelligence,* vol. 1, no. 5, pp. 206-215, 2019.

[3]  M. Ribeiro, S. Singh and C. Guestrin, "Why Should I Trust You?": Explaining the Predictions of Any Classifier," in *In Proceedings of the 22nd ACM SIGKDD international conference on knowledge discovery and data mining*, 2016.

[4]  M. Scott and L. Su-In, "A Unified Approach to Interpreting Model Predictions," *Advances in neural information processing systems,* vol. 30, pp. 4765-4774, 2017.

[5]  L. Gilpin , D. Bau and et al, "Explaining Explanations: An Overview of Interpretability of Machine Learning," in *In 2018 IEEE 5th International Conference on data science and advanced analytics (DSAA)*, 2018.

[6]  M. Frasca, D. La Torre and et al, "Explainable and interpretable artificial intelligence in medicine: a systematic bibliometric review," *Discover Artificial Intelligence,* vol. 4, no. 1, p. 15, 2024.

[7]  B. de Vries, G. Zwezerijnen and et al, "Explainable artificial intelligence (XAI) in radiology and nuclear medicine: a literature review," *Frontiers in Medicine,* vol. 10, p. 1180773, 2023.

[8]  P. Korica, N. Gayar and W. Pang, "Explainable Artificial Intelligence in Healthcare: Opportunities, Gaps and Challenges and a Novel Way to Look at the Problem Space," in *In International Conference on Intelligent Data Engineering and Automated Learning*, 2021.

[9]  A. Holzinger, C. Biemann and et al, "What do we need to build explainable AI systems?," *arXiv preprint arXiv:1712.09923,* 2017.

[10] A. Arrieta, N. Díaz-Rodríguez and et al, "Explainable Artificial Intelligence (XAI): : Concepts, taxonomies, opportunities and challenges toward responsible AI," *Information Fusion,* vol. 58, pp. 82-115, 2020.

[11] S. Band, A. Yarahmadi and et al, "Application of explainable artificial intelligence in medical health: A systematic review of interpretability methods," *Informatics in Medicine Unlocked,* vol. 40, p. 101286, 2023.

[12] M. Tomaszewski and R. Gillies, "The Biological Meaning of Radiomic Features," *RSNA, Radiology,* vol. 289, no. 3, pp. 505-516, 2021.

[13] P. Lambin, E. Rios-Velazquez and et al, "Radiomics: Extracting more information from medical images using advanced feature analysis," *European Journal of Cancer,* vol. 48, no. 4, pp. 441-446, 2012.

[14] A. Aerts, E. Velazquez and et al, "Decoding tumour phenotype by noninvasive imaging using a quantitative radiomics approach," *Nature Communications,* vol. 5, no. 1, p. 4006, 2014.

[15] M. Avanzo, J. Stancanello and I. El Naqa, "Beyond imaging: The promise of radiomics," *Physica Medica,* vol. 38, pp. 122-139, 2017.

[16] A. Zwanenburg , M. Vallières and et al, "The image biomarker standardization initiative: Standardized quantitative radiomics for high-throughput image-based phenotyping," *Radiology,* vol. 295, no. 2, pp. 328-338, 5 2020.

[17] S. Yip and H. Aerts , "Applications and limitations of radiomics," *Physics in Medicine & Biology,* vol. 61, no. 13, p. R150, 2016.

[18] Y. Huang, Z. Liu and et al, "Radiomics Signature: A Potential Biomarker for the Prediction of Disease-Free Survival in Early-Stage (I or II) Non—Small Cell Lung Cancer," *Radiology,* vol. 281, no. 3, pp. 947-957, 2016.

[19] V. Kumar, Y. Gu and et al, "Radiomics: the process and the challenges," *Magnetic Resonance Imaging,* vol. 30, no. 9, pp. 1234-1248, 2012.

[20] M. Hosseinzadeh, A. Gorji and et al, "Prediction of Cognitive Decline in Parkinson's Disease Using Clinical and DAT SPECT Imaging Features, and Hybrid Machine Learning Systems," *MDPI,* vol. 13, no. 10, p. 1691, 2023.

[21] M. Salmanpour, M. Shamsaei and A. Rahmim, "Feature selection and machine learning methods for optimal identification and prediction of subtypes in Parkinson's disease," *Computer Methods and Programs in Biomedicine,* vol. 206, p. 106131, 2021.

[22] M. Salmanpour, M. Hosseinzadeh and et al, "Advanced survival prediction in head and neck cancer using hybrid machine learning systems and radiomics features," in *In Medical Imaging 2022: Biomedical Applications in Molecular, Structural, and Functional Imaging*, 2022.

[23] J. Weinreb, J. Barentsz and e. al, "PI-RADS Prostate Imaging - Reporting and Data System: 2015, Version 2," *European Urology,* vol. 69, no. 1, pp. 16-40, 2016.

[24] B. Turkbey, A. Rosenkrantz and et al, "Prostate Imaging Reporting and Data System Version 2.1: 2019 Update of Prostate Imaging Reporting and Data System Version 2," *European Urology,* vol. 76, no. 3, pp. 340-351, 2019.

[25] A. Rosenkrantz, L. Ginocchio and et al, "Interobserver Reproducibility of the PI-RADS Version 2 Lexicon: A Multicenter Study of Six Experienced Prostate Radiologists," *Radiology,* vol. 280, no. 3, pp. 793-804, 2016.

[26] M. Salmanpour, S. Rezaeijo and et al, "Deep versus Handcrafted Tensor Radiomics Features: Prediction of Survival in Head and Neck Cancer Using Machine Learning and Fusion Techniques," *Diagnostics,* vol. 13, no. 10, p. 1696, 2023.

[27] B. Zhao, "Understanding Sources of Variation to Improve the Reproducibility of Radiomics," *Frontiers in Oncology,* vol. 11, p. 633176, 2021.

[28] J. Barentsz, J. Richenberg and et al, "ESUR prostate MR guidelines 2012," *European Radiology,* vol. 22, pp. 746-757, 2012.

[29] M. Christoph, Interpretable machine learning: A guide for making black box models explainable, Leanpub, 2020.

[30] A. Padhani, J. Barentsz and et al, "PI-RADS Steering Committee: The PI-RADS Multiparametric MRI and MRI-directed Biopsy Pathway," *RSNA, Radiology,* vol. 292, no. 2, pp. 464-474, 2019.

[31] X. Fave , L. Zhang and et al, "Delta-radiomics features for the prediction of patient outcomes in non–small cell lung cancer," *Scientific Reports,* vol. 7, no. 1, p. 588, 2017.

[32] G. Varriano, P. Guerriero and et al, "Explainability of radiomics through formal methods," *Computer Computer Computer Computer Computer Computer,* vol. 220, p. 106824, 2022.

[33] R. Cuocolo, A. Stanzione and et al, "Clinically significant prostate cancer detection on MRI: A radiomic shape," *European Journal of Radiology,* vol. 116, pp. 144-149, 2019.

[34] Y. Huang, C. Liang and et al, "Development and Validation of a Radiomics Nomogram for Preoperative Prediction of Lymph Node Metastasis in Colorectal Cancer," *Journal of Clinical Oncology,* vol. 34, no. 18, pp. 2157-2164, 2016.

[35] C. Parmar, P. Grossmann and et al, "Radiomic Machine-Learning Classifiers for Prognostic Biomarkers of Head and Neck Cancer," *Frontiers in oncology,* vol. 5, p. 272, 2015.

[36] T. Coroller, V. Agrawal and et al, "Radiomic phenotype features predict pathological response in non-small cell lung cancer," *Radiotherapy and Oncology,* vol. 119, no. 3, pp. 480-486, 2016.

[37] M. Salmanpour, M. Shamsaei and et al, "Machine learning methods for optimal prediction of outcome in Parkinson's disease," in *2018 IEEE Nuclear Science Symposium and Medical Imaging Conference Proceedings (NSS/MIC)*, 2018.

[38] M. Salmanpour, M. Hosseinzadeh and et al, "Deep versus handcrafted tensor radiomics features: Application to survival prediction in head and neck cancer," in *EUROPEAN JOURNAL OF NUCLEAR MEDICINE AND MOLECULAR IMAGING*, 2022.

[39] M. Salmanpour, A. Saberi and et al, "Optimal feature selection and machine learning for prediction of outcome in Parkinson's disease," in *Journal of Nuclear Medicine*, 2020.

[40] M. Salmanpour, M. Shamsaei and et al, "Hybrid machine learning methods for robust identification of Parkinson's disease subtypes," in *Journal of Nuclear Medicine*, 2020.

[41] R. Gillies, P. Kinahan and H. Hricak , "Radiomics: Images Are More than Pictures, They Are Data," *Radiology,* vol. 278, no. 2, pp. 563-577, 2015.

[42] V. Liberini and et al, "Radiomics and artificial intelligence in prostate cancer: new tools for molecular hybrid imaging and theragnostics," *European Radiology Experimental,* 2022.

[43] D. Truhn, S. Schrading and et al, "Radiomic versus Convolutional Neural Networks Analysis for Classification of Contrast-enhancing Lesions at Multiparametric Breast MRI," *RSNA, Radiology,* vol. 290, no. 2, pp. 290-297, 2019.

[44] M. Salmanpour, M. Shamsaei and et al, "Hybrid machine learning methods for robust identification of Parkinson's disease subtypes," in *Journal of Nuclear Medicine*, 2020.

[45] J. Bibault, L. Xing and et al, "Radiomics: A primer for the radiation oncologist," *Cancer/Radiothérapie,* vol. 24, no. 5, pp. 403-410, 2020.

[46] S. Konate, "Interpretability of machine learning models for medical image analysis," *PhD thesis, Queensland University of Technology.,* 2024.

[47] J. Mlynář and et al, "Making sense of radiomics: insights on human–AI collaboration in medical interaction from an observational user study," *Frontiers in Communication,* 2024.

[48] J. Mi, A. Li and L. Zhou, "Review study of interpretation methods for future interpretable machine learning," *IEEE Access,* vol. 8, pp. 191969-191985, 2020.

[49] S. Dash, O. Gunluk and D. Wei, "Boolean Decision Rules via Column Generation," in *32nd Conference on Neural Information Processing Systems*, 2018.

[50] F. Yang, K. He and et al, "Learning interpretable decision rule sets: A submodular optimization approach," in *in Proceedings of the 35th International Conference on Neural Information Processing Systems*, 2021.

[51] J. Yu,, A. Ignatiev, and et al, "Learning optimal decision sets and lists with sat," *Journal of Artificial Intelligence Research,* vol. 72, p. 1251–1279, 2021.

[52] G. Zhang and A. Gionis,, "Diverse rule sets," in *in Proceedings of the 26th ACM SIGKDD International Conference on Knowledge Discovery and Data Mining*, 2020.

[53] Y. Wu, L. Zhang and et al, "Interpretable machine learning for personalized medical recommendations: A LIME-based approach," *Diagnostic,* vol. 13, no. 16, p. 2681, 2023.

[54] M. Virgolin, D. Lorenzo and et al, "Model learning with personalized interpretability estimation (ML-PIE)," in *In Proceedings of the Genetic and Evolutionary Computation Conference Companion*, 2021.

[55] F. Valente, S. Paredes and et al, "Interpretability, personalization and reliability of a machine learning based clinical decision support system," *Data Mining and Knowledge Discovery,* vol. 36, no. 3, pp. 1140-1173, 2022.

[56] T. Wang, C. Rudin and et al, "A Bayesian framework for learning rule sets for interpretable classification," *Journal of Machine Learning Research,* vol. 18, no. 70, pp. 1-37, 2017.

[57] J. Kelleher , B. Mac Namee and A. D'arcy, Fundamentals of machine learning for predictive data analytics: algorithms, worked examples, and case studies, MIT press, Cambridge, 2020.

[58] H. Almuallim and T. Dietterich, "Efficient algorithms for identifying relevant features," Oregon State University. Department of Computer Science, 1992.

[59] D. Theng and K. Bhoyar, "Feature selection techniques for machine learning: a survey of more than two decades of research," *Knowledge and Information Systems,* vol. 66, no. 3, pp. 1575-1637, 2024.

[60] H. Kumar, M. Shafiq and et al, "A Review on the Classification of Partial Discharges in Medium-Voltage Cables: Detection, Feature Extraction, Artificial Intelligence-Based Classification, and Optimization Techniques," *Energies,* vol. 17, no. 5, p. 1142, 2024.

[61] G. Hajianfar, S. Kalayinia and et al, "Prediction of Parkinson's disease pathogenic variants using hybrid Machine learning systems and radiomic features," *Physica Medica,* vol. 113, p. 102647, 2023.

[62] M. Salmanpour, M. Hosseinzadeh and et al, "Advanced survival prediction in head and neck cancer using hybrid machine learning systems and radiomics features," in *Medical Imaging 2022: Biomedical Applications in Molecular, Structural, and Functional Imaging*, 2022.

[63] M. Salmanpour, M. Hosseinzadeh and et al, "Cognitive Outcome Prediction in Parkinson's Disease Using Hybrid Machine Learning Systems and Radiomics Features," in *Journal of Nuclear Medicine*, 2022.

[64] M. Salmanpour, G. Hajianfar and et al, Advanced automatic segmentation of tumors and survival prediction in head and neck cancer, Springer International Publishing, 2021, pp. 202-210.

[65] M. Salmanpour, M. Shamsaei and et al, "Optimized machine learning methods for prediction of cognitive outcome in Parkinson's disease," *Computers in biology and medicine,* vol. 111, p. 103347, 2019.

[66] M. Salmanpour, M. Shamsaei and et al, "Machine learning methods for optimal prediction of motor outcome in Parkinson's disease," *Physica Medica,* vol. 69, pp. 233-240, 2020.

[67] E. Huang, J. O'Connor and et al, "Criteria for the translation of radiomics into clinically useful tests," *Nature reviews Clinical oncology,* vol. 20, no. 2, pp. 69-82, 2023.

[68] R. Stoyanova, M. Takhar and et al, "Prostate cancer radiomics and the promise of radiogenomics," *Translational cancer research,* vol. 5, no. 4, p. 432, 2016.

[69] K. Gnep, A. Fargeas and et al, "Haralick textural features on T2-weighted MRI are associated with biochemical recurrence following radiotherapy for peripheral zone prostate cancer," *Journal of Magnetic Resonance Imaging,* vol. 45, no. 1, pp. 103-17, 2017.

[70] P. Kickingereder, S. Burth and et al, "Radiomic Profiling of Glioblastoma: Identifying an Imaging Predictor of Patient Survival with Improved Performance over Established Clinical and Radiologic Risk Models," *RSNA, Radiology,* vol. 280, no. 3, pp. 880-889, 2016.

[71] M. Cooperberg, D. Pasta and et al, "The UCSF Cancer of the Prostate Risk Assessment (CAPRA) Score: a straightforward and reliable preoperative predictor of disease recurrence after radical prostatectomy," *The Journal of urology,* vol. 173, no. 9, p. 1938, 2005.

[72] N. Papanikolaou, C. Matos and D. Koh, "How to develop a meaningful radiomic signature for clinical use in oncologic patients," *Cancer Imaging,* vol. 20, no. 1, p. 33, 2020.

[73] F. Mercaldo and et al, "An Explainable Method for Lung Cancer Detection and Localisation from Tissue Images through Convolutional Neural Networks," *MDPI Open Access Journals,* 2024.

[74] V. Vimbi, N. Shaffi and M. Mahmud, "Interpreting artificial intelligence models: a systematic review on the application of LIME and SHAP in Alzheimer's disease detection," *Brain Informatics,* vol. 11, no. 1, p. 10, 2024.

[75] J. Van Timmeren, D. Cester and et al, "Radiomics in medical imaging—"how-to" guide and critical reflection," *Insights into Imaging,* vol. 11, no. 1, p. 91, 2020.

[76] Y. Brima and M. Atemkeng, "Saliency-driven explainable deep learning in medical imaging: bridging visual explainability and statistical quantitative analysis," *BioData Mining,* vol. 17, no. 1, pp. 1-33, 2024.

[77] S. Lundberg, G. Erion and et al, "From Local Explanations to Global Understanding with Explainable AI for Trees," *Nature Machine Intelligence,* vol. 2, no. 1, pp. 56-67, 2020.

[78] A. Singh, S. Sengupta and V. Lakshminarayanan, "Explainable Deep Learning Models in Medical Image Analysis," *Journal of Imaging,* vol. 6, no. 6, p. 52, 2020.

[79] J. Van Griethuysen, A. Fedorov and et al, "Computational Radiomics System to Decode the Radiographic Phenotype," *Cancer Research,* vol. 77, no. 21, pp. 104-7, 2017.

[80] L. Kolla and R. B. Parikh, "Uses and limitations of artificial intelligence for oncology," *American Cancer Society,* 2024.

[81] S. Duque Anton, D. Schneider and H. Schotten, "On Explainability in AI-Solutions: A Cross-Domain Survey," in *In International Conference on Computer Safety, Reliability, and Security*, 2022.

[82] S. Natarajan, A. Priester, D. Margolis, J. Huang and L. Marks, "Prostate MRI and Ultrasound With Pathology and Coordinates of Tracked Biopsy (Prostate-MRI-US-Biopsy) (version 2) [Data set]," *Cancer Imaging Arch,* vol. 10, p. 7937, 2020.

[83] J. Hu, E. Chang and et al, "Targeted prostate biopsy to select men for active surveillance: do the Epstein criteria still apply?," *The Journal of urology,* vol. 192, no. 2, pp. 385-390, 2014.

[84] M. Litwin, R. Hays, A. Fink, P. Ganz, B. Leake and R. Brook, "The UCLA Prostate Cancer Index: development, reliability, and validity of a health-related quality of life measure," *Medical care,* vol. 36, no. 7, pp. 1002-1012, 1998.

[85] G. Stiglic, P. Kocbek and et al, "Interpretability of machine learning-based prediction models in healthcare," *Wiley Interdisciplinary Reviews: Data Mining and Knowledge Discovery,* vol. 10, no. 5, p. e1379, 2020.

[86] P. Linardatos, V. Papastefanopoulos and S. Kotsiantis, "Explainable AI: A Review of Machine Learning Interpretability Methods," *Entropy,* vol. 23, no. 1, p. 18, 2020.

[87] D. Carvalho, E. Pereira and J. Cardoso, "Machine Learning Interpretability: A Survey on Methods and Metrics," *Electronics,* vol. 8, no. 8, p. 832, 2019.

[88] S. Martin, F. Townend and et al, "Interpretable machine learning for dementia: a systematic review," *Alzheimer's & Dementia,* vol. 19, no. 5, pp. 2135-2149, 2023.

[89] C. Azodi, J. Tang and S. Shiu, "Opening the black box: interpretable machine learning for geneticists," *Trends in genetics,* vol. 36, no. 6, pp. 442-455, 2020.

[90] G. Allen, L. Gan and L. Gan, "Interpretable machine learning for discovery: Statistical challenges and opportunities," *Annual Review of Statistics and Its Application,* vol. 11, pp. 97-121, 2023.

[91] A. Jović, K. Brkić and N. Bogunović, "A review of feature selection methods with applications," in *38th International Convention on Information and Communication Technology, Electronics and Microelectronics (MIPRO)*, 2015.

[92] A. Bommert, X. Sun and et al, "Benchmark for filter methods for feature selection in high-dimensional classification data," *Computational Statistics & Data Analysis,* vol. 143, p. 106839, 2020.

[93] F. Emmert-Streib and M. Dehmer, "High-Dimensional LASSO-Based Computational Regression Models: Regularization, Shrinkage, and Selection," *Machine Learning and Knowledge Extraction,* vol. 1, no. 1, pp. 359-383, 2019.

[94] W. Kanyongo and A. Ezugwu, "Feature selection and importance of predictors of non-communicable diseases medication adherence from machine learning research perspectives," *Informatics in Medicine Unlocked,* vol. 38, p. 101232, 2023.

[95] S. Kost, O. Rheinbach and H. Schaeben, "Using logistic regression model selection towards interpretable machine learning in mineral prospectivity modeling," *Geochemistry,* vol. 81, no. 4, p. 125826, 2021.

[96] Y. Chen , D. Wang and et al, "Gaussian assumptions-free interpretable linear discriminant analysis for locating informative frequency bands for machine condition monitoring," *Mechanical Systems and Signal Processing,* vol. 199, p. 110492, 2023.

[97] M. Salmanpour, M. Hosseinzadeh and et al, "Prediction of TNM stage in head and neck cancer using hybrid machine learning systems and radiomics features," vol. 12033, pp. 662-667, 2022.

[98] A. Fathi Jouzdani, A. Abootorabi and et al, "Impact of Clinical Features Combined with PET/CT Imaging Features on Survival Prediction of Outcome in Lung Cancer," *Journal of Nuclear Medicine,* vol. 65, no. supplement 2, pp. 242130-242130, 2024.

[99] M. Salmanpour, G. Hajianfar and et al, "Multitask outcome prediction using hybrid machine learning and PET-CT fusion radiomics," vol. 62, pp. 1424-1424, 2021.

[100] M. Salmanpour, A. Saberi and et al, "Optimal feature selection and machine learning for prediction of outcome in Parkinson's disease," vol. 61, pp. 524-524, 2020.

[101] N. Rodrigues, J. Almeida and et al, "Deep Learning Features Can Improve Radiomics-Based Prostate Cancer Aggressiveness Prediction," *JCO Clinical Cancer Informatics,* vol. 8, p. e2300180, 2024.

[102] J. Huang, C. He and et al, "Development and validation of a clinical-radiomics model for prediction of prostate cancer: a multicenter study," *World Journal of Urology,* vol. 42, no. 1, p. p.275, 2024.

[103] C. Chen, Z. Yang and et al, "Prostate heterogeneity correlates with clinical features on multiparametric MRI," *Abdominal Radiology,* vol. 46, pp. 5369-5376, 2021.

[104] M. Echevarria, A. O. Naghavi and et al, "MRI Heterogeneity in Prostate Cancer Predicts for Aggressive Pathology," *International Journal of Radiation Oncology,* vol. 99, no. 2, p. E229, 2017.

[105] N. Raison, P. Servian and et al, "Is tumour volume an independent predictor of outcome after radical prostatectomy for high-risk prostate cancer?," *Prostate cancer and prostatic diseases,* vol. 26, no. 2, pp. 282-286., 2023.

[106] M. Salmanpour, I. Shiri and et al, "ViSERA: Visualized & Standardized Environment for Radiomics Analysis-A Shareable, Executable, and Reproducible Workflow Generator," in *2023 IEEE Nuclear Science Symposium, Medical Imaging Conference and International Symposium on Room-Temperature Semiconductor Detectors (NSS MIC RTSD)*, Vancouver, 2023.